\documentclass[a4paper,12pt]{article}
\usepackage{feynmp-auto}

\usepackage{hepnicenames}
\usepackage{enumerate}

\usepackage{mathrsfs,graphicx,rotating,amsmath,amsfonts,mathtools,booktabs,amssymb,wasysym}
\usepackage{hyperref}\usepackage{slashed}
\usepackage[nosort]{cite}
\usepackage[table,xcdraw,dvipsnames]{xcolor}
\usepackage{youngtab}
\usepackage{bm}
\usepackage{graphicx}
\usepackage{multirow,multicol}
\hypersetup{colorlinks,bookmarksopen,bookmarksnumbered,
linkcolor=blus,pdfstartview=FitH,urlcolor=rossos,citecolor=verde}
\allowdisplaybreaks

\renewcommand{\S}{{\cal S}}

\newcommand{\Mtexp}{172.5\pm 0.5\GeV}

\newcommand{\LQCD}{\Lambda_{\rm QCD}}
\newcommand{\mio}[1]{}

 \newcommand{\med}[1]{\langle #1\rangle}

\def\bpm{\begin{pmatrix}}
\def\epm{\end{pmatrix}}
\def\mcal{\mathscr}
\usepackage{mathrsfs}

 \newcommand{\fig}[1]{~\ref{fig:#1}}
\newcommand{\sfrac}[2]{#1/#2}

\allowdisplaybreaks
\usepackage{multicol}
\usepackage{color}
\definecolor{rosso}{cmyk}{0,1,1,0.4}
\definecolor{rossos}{cmyk}{0,1,1,0.55}
\definecolor{rossoc}{cmyk}{0,1,1,0.2}
\definecolor{blu}{cmyk}{1,1,0,0.3}
\definecolor{blus}{cmyk}{1,1,0,0.6}
\definecolor{bluc}{cmyk}{1,1,0,0.1}
\definecolor{verde}{cmyk}{0.92,0,0.59,0.25}
\definecolor{verdec}{cmyk}{0.92,0,0.59,0.15}
\definecolor{verdes}{cmyk}{0.92,0,0.59,0.4}

\newcommand{\eq}[1]{~{\rm (\ref{eq:#1})}}

\newcommand{\MeV}{\,{\rm MeV}}
\newcommand{\GeV}{\,{\rm GeV}}
\newcommand{\TeV}{\,{\rm TeV}}
\newcommand{\cm}{\,{\rm cm}}

\def\circa#1{\,\raise.3ex\hbox{$#1$\kern-.75em\lower1ex\hbox{$\sim$}}\,}

\newcommand{\beq}{\begin{equation}}
\newcommand{\eeq}{\end{equation}}
\newcommand{\mb}[1]{\mbox{\boldmath $#1$}}

\newcommand{\bea}{\begin{eqnarray}}
\newcommand{\eea}{\end{eqnarray}}
\newcommand{\be}{\begin{equation}}
\newcommand{\ee}{\end{equation}}
\font\tenrsfs=rsfs10 at 12pt
\font\sevenrsfs=rsfs7
\font\fiversfs=rsfs5
\newfam\rsfsfam
\textfont\rsfsfam=\tenrsfs
\scriptfont\rsfsfam=\sevenrsfs
\scriptscriptfont\rsfsfam=\fiversfs

\newsavebox\MBox

\def\circa#1{\,\raise.3ex\hbox{$#1$\kern-.75em\lower1ex\hbox{$\sim$}}\,}
\makeatletter

\font\ital=cmu10 

\def\hhref#1{\href{http://arxiv.org/abs/#1}{arXiv:#1}}
\usepackage{xstring} 
\newcommand{\hhrefq}[1]{\IfSubStr{#1}{:}{\href{http://inspirehep.net/search?ln=en&ln=en&p=#1&of=hb&action_search=Search&sf=&so=d&rm=&rg=25&sc=0}{InSpires:#1}}{\hhref{#1}}}

\def\art{\@ifnextchar[{\eart}{\oart}}
\def\eart[#1]#2#3#4#5#6{{\rm #2}, {\em #3 \bf #4} {\rm (#6) #5} ({\em #1})}
\def\article{\@ifnextchar[{\earticle}{\oarticle}}
\def\oarticle#1#2#3#4#5#6{{\rm #1}, {\ital ``#6''}, {\rm #2 #3 (#5) #4}}
\def\earticle[#1]#2#3#4#5#6#7{{\rm #2}, {\ital ``#7''}, {\rm #3 #4 (#6) #5}  [\hhrefq{#1}]}
\def\hepart[#1]#2{{\rm #2, \sl#1}}
\def\heparticle[#1]#2#3{#2, {\ital ``#3''} [\hhrefq{#1}]}
\newcommand{\doi}[1]{\href{http://dx.doi.org/#1}{[link]}}

\renewenvironment{thebibliography}[1]
     {\begin{multicols}{2}[\section*{\refname}]%
      \@mkboth{\MakeUppercase\refname}{\MakeUppercase\refname}%
      \list{\@biblabel{\@arabic\c@enumiv}}%
           {\settowidth\labelwidth{\@biblabel{#1}}%
            \leftmargin\labelwidth
            \advance\leftmargin\labelsep
            \@openbib@code
            \usecounter{enumiv}%
            \let\p@enumiv\@empty
            \renewcommand\theenumiv{\@arabic\c@enumiv}}%
      \sloppy
      \clubpenalty4000
      \@clubpenalty \clubpenalty
      \widowpenalty4000%
      \sfcode`\.\@m}
     {\def\@noitemerr
       {\@latex@warning{Empty `thebibliography' environment}}%
      \endlist\end{multicols}}

\newcounter{alphaequation}[equation]
\def\thealphaequation{\theequation\hbox to
0.6em{\hfil\alph{alphaequation}\hfil}}
\def\eqnsystem#1{
\def\@eqnnum{{\rm (\thealphaequation)}}
\def\@@eqncr{\let\@tempa\relax \ifcase\@eqcnt \def\@tempa{& & &} \or
  \def\@tempa{& &}\or \def\@tempa{&}\fi\@tempa
  \if@eqnsw\@eqnnum\refstepcounter{alphaequation}\fi
\global\@eqnswtrue\global\@eqcnt=0\cr}
\refstepcounter{equation} \let\@currentlabel\theequation \def\@tempb{#1}
\ifx\@tempb\empty\else\label{#1}\fi
\refstepcounter{alphaequation}
\let\@currentlabel\thealphaequation
\global\@eqnswtrue\global\@eqcnt=0 \tabskip\@centering\let\\=\@eqncr
$$\halign to \displaywidth\bgroup \@eqnsel\hskip\@centering
$\displaystyle\tabskip\z@{##}$&\global\@eqcnt\@ne
\hskip2\arraycolsep\hfil${##}$\hfil& \global\@eqcnt\tw@\hskip2\arraycolsep
$\displaystyle\tabskip\z@{##}$\hfil
\tabskip\@centering&\llap{##}\tabskip\z@\cr}
\def\endeqnsystem{\@@eqncr\egroup$$\global\@ignoretrue} \makeatother

\definecolor{Gray}{gray}{0.95}

\def\bal#1\eal{\begin{align}#1\end{align}}

\begin{document}

{CERN-TH-2018-065\hfill IFUP-TH/2018}

\vspace{2cm}

\begin{center}
{\Large\LARGE \bf \color{rossos}
Dark Matter in the Standard Model?}\\[1cm]
{\bf Christian Gross}$^{a}$,
{\bf Antonello Polosa}$^{b}$, \\
{\bf Alessandro Strumia}$^{a,c}$,
{\bf Alfredo Urbano}$^{d,c}$,
{\bf Wei Xue}$^{c}$
\\[7mm]

{\it $^a$ Dipartimento di Fisica dell'Universit{\`a} di Pisa and INFN, Sezione di Pisa, Italy}\\[1mm]
{\it $^b$ Dipartimento di Fisica e INFN, Sapienza Universit\`a di Roma, I-00185, Roma, Italy}\\[1mm]
{\it $^c$ Theoretical Physics Department, CERN, Geneva, Switzerland}\\[1mm]
{\it $^d$ INFN, Sezione di Trieste, SISSA, via Bonomea 265, 34136 Trieste, Italy}

\vspace{0.5cm}

{\large\bf\color{blus} Abstract}
\begin{quote}
\end{quote}

\thispagestyle{empty}
\bigskip

\end{center}
\begin{quote}
\large\noindent\color{blus} 
We critically reexamine two possible Dark Matter candidates within the Standard Model.
First, we consider the $uuddss$ hexa-quark.
Its QCD binding energy could be large enough to make it (quasi) stable.
We show that the cosmological Dark Matter abundance is reproduced thermally if its mass is $1.2\GeV$.
However, we also find that such mass is excluded by stability of Oxygen nuclei.
Second, we consider the possibility that the instability in the Higgs potential
leads to the formation of primordial black holes while avoiding vacuum decay during inflation.
We show that the non-minimal Higgs coupling to gravity must be as small as allowed by quantum corrections,
$|\xi_H| < 0.01$.
Even so, one must assume that the Universe survived 
in $e^{120}$ independent regions to
fluctuations that lead to vacuum decay with probability 1/2 each.
\end{quote}

\newpage
\tableofcontents

\setcounter{footnote}{0}

\section{Introduction}
In this work we critically re-examine two different intriguing possibilities that challenge 
the belief that the existence of Dark Matter (DM) implies new physics beyond the Standard Model (SM).

\subsubsection*{DM as the $uuddss$ hexa-quark}
The binding energy of the
hexa-quark di-baryon $uuddss$ is expected to be large, 
given that the presence of the strange quark $s$ allows it to be a scalar, isospin singlet~\cite{neutronstar},
called $H$ or $\S$, and sometimes named $s$exa-quark.
A large binding energy might make $\S$ light enough that it is stable or long lived.
All possible decay modes of a free $\S$ are kinematically forbidden if $\S$ is lighter
than about $1.87\GeV$.
Then $\S$ could be a  Dark Matter candidate within the Standard Model~\cite{Farrar:2003qy,1708.08951,1711.10971}.

In section~\ref{mass} we use the recent theoretical and experimental 
progress about tetra- and penta-quarks to infer the mass of the $\S$ hexa-quark.
In section~\ref{cosmo} we present the first cosmological computation of the relic $\S$ abundance,
finding that the desired value is reproduced for $M_\S \approx 1.2\GeV$.
In section~\ref{SK} we revisit the bound from nuclear stability ($NN \to \S X$ production within nuclei)
at the light of recent numerical computations of one key ingredient: the nuclear wave-function~\cite{Lonardoni:2017egu}, finding that $\S$ seems excluded.

\subsubsection*{DM as primordial black holes}
Primordial Black Holes (PBH) are hypothetical relics which can originate 
from gravitational  collapse of sufficiently large density fluctuations.
The formation of PBHs is not predicted by standard inflationary cosmology: 
the primordial inhomogeneities observed on large cosmological scales are too small.
PBH can arise in models with large inhomogeneities on small scales, $k \gg {\rm Mpc}^{-1}$.
PBH as DM candidates are subject to various constraints.
BH lighter than $ 6~10^{-17}\, M_\odot$ are excluded because of Hawking radiation.
BH heavier than $10^5 M_\odot$ are safely excluded.
In the intermediate region, a variety of bounds make the possibility that $\Omega_{\rm PBH}=\Omega_{\rm DM}$
problematic but maybe not excluded --- the issue is presently subject to an intense debate.
According to~\cite{1807.11495} DM as PHB with mass $M\sim 10^{-15}M_\odot$
are not excluded, as previously believed. And the HSC/Subaru microlensing constraint on PBH~\cite{Niikura:2017zjd} is 
partially in the wave optics region. This can invalidate its bound below $\sim 10^{-11} M_\odot$.

Many ad hoc models that can produce PBH as DM have been proposed.
Recently~\cite{Espinosa:2017sgp} claimed that a mechanism of this type is present within the Standard Model
given that, for present best-fit values of the measured SM parameters,
the SM Higgs potential is unstable at $h > h_{\rm max} \sim 10^{10}\GeV$~\cite{1307.3536}.
We here critically re-examine the viability of the proposed mechanism, which assumes that
the Higgs, at some point during inflation,
has a homogeneous vev mildly above the top of the barrier
and starts rolling down.
When inflation ends, reheating adds a large thermal mass to the effective Higgs potential,
which, under certain conditions, brings the Higgs back to the origin, $h=0$~\cite{1505.04825}.
If falling stops very close to the disaster, this process generates inhomogeneities
which lead to the formation of primordial black holes.
In section~\ref{DMBH}  we extend the computations of~\cite{Espinosa:2017sgp} adding
a non-vanishing non-minimal coupling $\xi_H$ of the Higgs to gravity, 
which is unavoidably generated by quantum effects~\cite{Freedman:1974gs}.
We find that $\xi_H$ must be as small as allowed by quantum effects.
Under the assumptions made~\cite{Espinosa:2017sgp} we reproduce their results;
however in section~\ref{Homo} we also find that such assumptions imply an extreme fine-tuning.

%
%
%

\medskip

The first mechanism is affected by the observed 
baryon asymmetry, but does not depend on the unknown physics
that generates the baryon asymmetry.
The second possibility depends on inflation, but the mechanism only depends 
on the (unknown) value of the Hubble constant during inflation.
In both cases the DM candidate is part of the SM.
Conclusions are given in section~\ref{conc}.

\section{DM as the $uuddss$ hexa-quark}\label{S}
The hexa-quark $\S=uuddss$ is stable if all its possible decay modes are kinematically closed:
\beq \S\to\left\{\begin{array}{ll}
{\rm d} e\bar\nu_e & M_\S <  M_{\rm d} + M_e = 1.8761\GeV\\
 ppe e \bar\nu_e\bar\nu_e & M_\S < 2(M_p+ M_e)=1.8775\GeV\\
 pe\bar\nu_e n & M_\S < M_p + M_e + M_n = 1.8783\GeV\\
 nn & M_\S < 2 M_n = 1.8791\GeV
 \end{array}\right.\eeq
A stable $\S$ is a possible DM candidate.
A too light $\S$ can make nuclei unstable.
Scanning over all stable nuclei, we find that none of them gets
destabilised by single $\S$ emission 
if $M_\S>1.874\GeV$, with $^6$Li  giving the potentially highest sensitivity to $M_\S$.



\subsection{Mass of the hexa-quark from a di-quark model}\label{mass}
We estimate the mass of the hexa-quark ${\cal S}$ viewing it as a neutral scalar di-baryon constituted by three spin zero di-quarks
\be
{\cal S}=\epsilon^{\alpha \beta\gamma}[ud]_{\alpha,s=0}[us]_{\beta,s=0}[ds]_{\gamma,s=0}
\ee
where $\alpha,\beta,\gamma$ are color indices. This is possible thanks to the strange $s$ quark, while
spin zero di-quarks of the kind $[uu],[dd],[ss]$ 
are forbidden by Fermi statistics because of antisymmetry in color and spin.
We assume the effective Hamiltonian for the  hexa-quark~\cite{1611.07920}
\begin{equation}
H=\sum_{i\neq j=\{u,d,s\}} (m_{ij}+ 2\kappa_{ij} \; \bm S_i\cdot \bm S_j)
\label{colorspin}
\end{equation}
where the $\kappa_{ij}$ are  effective couplings determined by the strong interactions at 
low energies, color factors, quark masses and wave-functions at the origin.  The  $m_{ij}$ are the masses of the di-quarks in ${\cal S}$ made of $i$ and $j$ constituent quarks~\cite{cqm}. $\bm S_i$ is the spin of $i$-th quark. Another important assumption, which is well  motivated by studies on tetra-quarks~\cite{1611.07920}, is that spin-spin interactions are essentially within di-quarks and zero outside, as if they were sufficiently separated in space. 

Considering di-quark masses to be additive in the constituent quark masses, and 
 taking $q$ and $s$ constituent quark masses from the baryons
one finds
\bea
&& m_{[qq]}\simeq 0.72~\mathrm{GeV},\qquad m_{[qs]}\sim 0.90~\mathrm{GeV}\qquad
q=\{u,d\}.
\label{diqmass}
\eea
The chromomagnetic couplings $\kappa_{ij}$ could as well be derived in the constituent quark model using data 
on baryons 
\be
\kappa_{qq}\simeq 0.10~\mathrm{GeV},\qquad \kappa_{qs}\simeq 0.06~\mathrm{GeV}.
\ee
However it is known that to reproduce the masses of light scalar mesons, interpreted as tetraquarks, $\sigma(500), f_0(980), a_0(980), \kappa$~\cite{hep-ph/0407017}, we need
\be
\kappa_{qq}\simeq 0.33~\mathrm{GeV},\qquad \kappa_{qs}\simeq 0.27~\mathrm{GeV}.
\label{kmes}
\ee
Spin-spin  couplings in tetra-quarks are found to be about a factor of four larger compared to the spin-spin couplings among the same pairs of quarks in the baryons, which make also di-quarks. It is difficult to assess if this would change within an hexa-quark. At any rate we can attempt a simple  mass formula for ${\cal S}$
\be
M_{\cal S}=(m_{[qq]} -3/2\kappa_{qq})+2(m_{[qs]} -3/2\kappa_{qs})
\label{stima}
\ee
which in terms of light tetra-quark masses means $M_{\cal S}=M_{\sigma}/2+M_{f_0}$.
Using the determination of chromo-magnetic couplings from baryons we would obtain 
\be
M_{\cal S}\approx 2.17~\mathrm{GeV}
\label{msbar}
\ee
whereas keeping the chromo-magnetic couplings needed to fit light tetra-quarks gives
\be
M_{\cal S}\approx 1.2~\mathrm{GeV}
\ee
if the same values for the chromo-magnetic couplings to fit light tetra-quark masses are taken (or 1.4~GeV using~\eqref{kmes}). There is quite a lot of experimental information on tetra-quarks~\cite{1611.07920}, whereas hexa-quarks, for the moment, are purely hypothetical objects.  On purely qualitative grounds we might expect that the mass of ${\cal S}$ could be closer to the heavier value  being a di-baryon and not a di-meson (tetra-quark) like light scalar mesons.

In the absence of any other experimental information it is  impossible to provide an estimate of the theoretical uncertainty on~$M_\S$.

Lattice computations performed at unphysical values of quark masses
find small values for the $\S$ binding energy, about 13, 75, 20 MeV~\cite{1109.2889,1206.5219,1805.03966}.
Extrapolations to physical quark masses suggest that $\S$ does not have a sizeable binding energy, see e.g.~\cite{Shanahan:2011su}.
Furthermore, the  binding energy of the deuteron is small, indirectly disfavouring
a very large binding energy for the (somehow similar) $\S$,
which might too be a molecule-like state.\footnote{We thank M.~Karliner, A.~Francis, J.~Green for discussions.}
Despite of this, we over-optimistically treat $M_\S$ as a free parameter in the following.

We also notice that the ${\cal S}$ particle could be  much larger than what envisaged in~\cite{Farrar:2003qy,1708.08951,1711.10971}  and that its  coupling to photons, in the case of $2-3$~fm size (see the considerations on diquark-diquark repulsion at small distances in~\cite{Selem:2006nd}), could be relevant for momentum transfers $k$ as small as $k\sim 60$~MeV, compared to $k\approx 0.5$~GeV considered by Farrar.  

\subsection{Cosmological relic density of the hexa-quark}\label{cosmo}
We here compute the relic density of $\S$  Dark Matter, studying if it can match the measured value $\Omega_{\rm DM} h^2=0.1186$, i.e.~$ \Omega_{\rm DM}  \approx 5.3 \ \Omega_b$~\cite{Ade:2015xua}.
A key ingredient of the computation is the baryon asymmetry. Its value measured in CMB and BBN is
$ Y_B=\sfrac{n_B}{s}=0.8 ~10^{-10}$.
The DM abundance is reproduced (using $M_\S=1.2$~GeV for definiteness) for 
\beq
\frac{Y_\S}{Y_B}=\frac{\Omega_{\rm DM}}{\Omega_{B}} \frac{M_p}{M_\S} \approx 4.2.\eeq
Thereby the baryon asymmetry before $\S$ decoupling must be
\beq \label{eq:YBS}Y_{B\S}=Y_B + 2 Y_\S\approx 9.3 Y_B.\eeq

\begin{figure}[t]
\begin{center}
\includegraphics[width=.47\textwidth,trim=0 6pt 0 0]{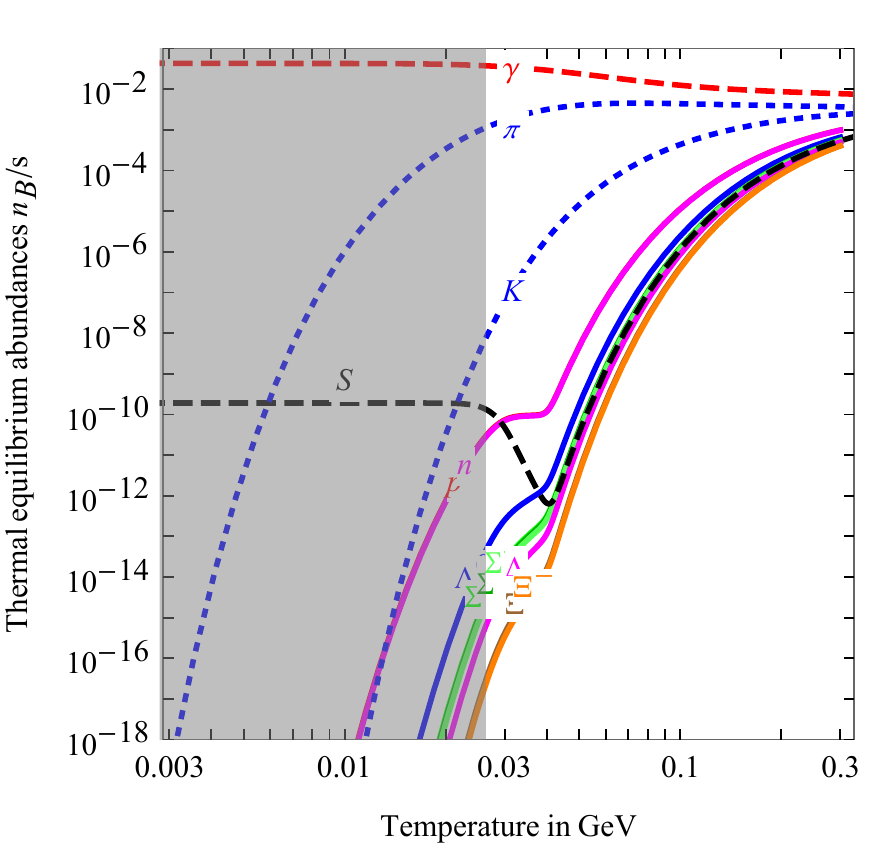}\qquad
\includegraphics[width=.45\textwidth]{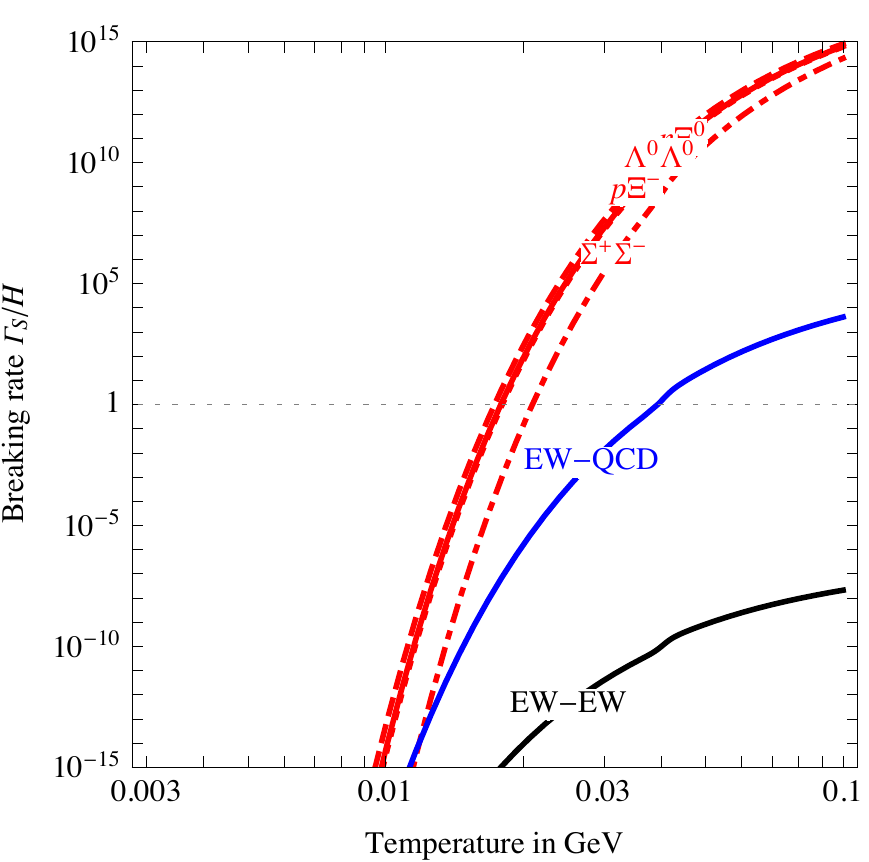}
\caption{\em \label{fig:baryonevo} {\bf Left}: thermal equilibrium values of hadron abundances
assuming $Y_{B\S} = 7.6~10^{-10}$ and $M_\S = 1.2\GeV$: the desired abundances are obtained
if decoupling of $\S$ interactions occurs at $T \sim 25\MeV$.
We see that in this phase anti-particles are negligible.
{\bf Right}: thermal equilibrium values of $\Gamma_\S/H$.}
\end{center}
\end{figure}

One needs to evolve a network of Boltzmann equations for the main hadrons:
$p$, $n$, $\Lambda$, $\Sigma^0$, $\Sigma^+$, $\Sigma^-$, $\Xi^0$, $\Xi^-$ and $\S$.
Strange baryons undergo weak decays with lifetimes $\tau \sim 10^{-10}\,{\rm s}$,
a few orders of magnitude faster than the Hubble time.
This means that such baryons stay in thermal equilibrium.
We thereby first compute the thermal equilibrium values taking into account the baryon asymmetry.
Thermal equilibrium implies that the chemical potentials $\mu(T)$ satisfy
\beq \mu_b= \mu_S/2,\qquad b = \{p,n,\Lambda, \ldots\}.\eeq
Their overall values are determined imposing that the total baryon asymmetry equals
\beq \sum_b \frac{n_b^{\rm eq}- n_{\bar b}^{\rm eq}}{s} + 2 \frac{n_\S^{\rm eq} -n^{\rm eq}_{\bar\S}}{s} = Y_{B\S} .\eeq
The equilibrium values can be analytically computed
in Boltzmann approximation (which becomes exact in the non-relativistic limit)
\beq n_i^{\rm eq}= g_i \frac{M_i^2T}{2\pi^2} {\rm K}_2(\frac{M_i}{T}) e^{\pm \mu_i/T}
\label{muequilibrium}
\eeq
where the $+$ ($-$) holds for (anti)particles.
We then obtain the  abundances in thermal equilibrium plotted in  fig.\fig{baryonevo},
assuming $M_\S = 1.2\GeV$.
We see that the desired $\S$ abundance is reproduced if the interactions that form/destroy $\S$ 
decouple at $T_{\rm dec} \approx 25 \MeV$.
This temperature is so low that baryon anti-particles have negligible abundances, 
and computations can more simply be done neglecting anti-particles.\footnote{Let us consider, for example,  the process
$\Lambda +\Lambda \leftrightarrow \S+ X$
where $X$ denotes other SM particles that do not carry the baryon asymmetry, such as pions.
Thermal equilibrium of the above process implies
\beq 
\frac{n_{\S}}{n_{\Lambda}^2} =
 \frac{n^{\rm eq}_{S}}{n^{\rm eq 2}_{\Lambda}} =
 \frac{g_{\S}}{g_{\Lambda}^2} \left(\frac{2\pi M_\S }{M_\Lambda^2 T}\right)^{3/2} e^{\sfrac{(2 M_\Lambda-M_S)}{T}} .\eeq
Inserting $n_\Lambda \sim n_p \, e^{(M_p-M_\Lambda)/T}$ with $n_p \approx Y_{B\S} s$   gives 
 \beq
 \frac{n_\S}{n_p} \sim \frac{n_\Lambda^2}{n_p}  e^{(2 M_\Lambda-M_\S)/T} \sim Y_{B\S} e^{(2 M_{p}-M_{\S})/T} \eeq
Namely, $n_\S \ll n_p$ at large $T$; $n_\S\gg n_p$ at low $T$.
A DM abundance comparable to the baryon abundance is only obtained if  reactions
that form $\S$ decouple at the $T$ in eq.\eq{Tdec}.}
Then, the desired decoupling temperature is simply estimated imposing
$n^{\rm eq}_\S/n^{\rm eq}_p \sim Y_{B\S} e^{(2M_p - M_\S)/T} \sim 1$,
and decreases if $M_\S$ is heavier:
\beq \label{eq:Tdec}
T_{\rm dec}|_{\rm desired} \approx \frac{2M_p -M_\S}{|\ln Y_{B\S}|} \approx  89\MeV- 0.048 M_\S .\eeq

To compute the decoupling temperature, we consider the
three different kind of processes that can lead to formation of $\S$:
\begin{enumerate}

\item Strong interactions of two heavier QCD hadrons that contain the needed two $s$ quarks.
One example is $\Lambda \Lambda \leftrightarrow \S X$, where $X$ denotes pions.
These are doubly Boltzmann suppressed by $e^{-2m_s/T}$ at temperatures $T< m_s$.

\item Strong interactions of one heavier strange hadron and weak $\Delta S=1$ interactions
that form the other $s$ (as $d \bar u \to s\bar u$) from lighter hadrons.
One example is $ p \Lambda \leftrightarrow \S X$.
These are singly  Boltzmann suppressed by $e^{-m_s/T}$ and by $G_{\rm F}^2 \LQCD^4 \sim 10^{-10}$.

\item Double-weak interactions that form two $s$ quarks starting from lighter hadrons.
One example is $pp \to \S X$.
These are doubly suppressed by $(G_{\rm F}^2 \LQCD^4)^2$.
\end{enumerate}
At $T \sim 25 \MeV$ the abundance of strange hadrons is still large enough that QCD processes dominate
over EW processes: 
interactions that  form and destroy  $\S$ proceed dominantly through QCD collisions
of strange hadrons:
%
%
\beq \label{eq:Sprod}
\Lambda \Lambda, n \Xi^0, p\Xi^- , \Sigma^+ \Sigma^- \leftrightarrow \S X\eeq
where $X$ can be a $\pi^0$ or a $\gamma$,
as preferred by approximate isospin conservation.
The $\Lambda$ can be substituted by the $\Sigma^0$.

\medskip

 \begin{figure}[t]
\begin{center}
\includegraphics[width=.7\textwidth]{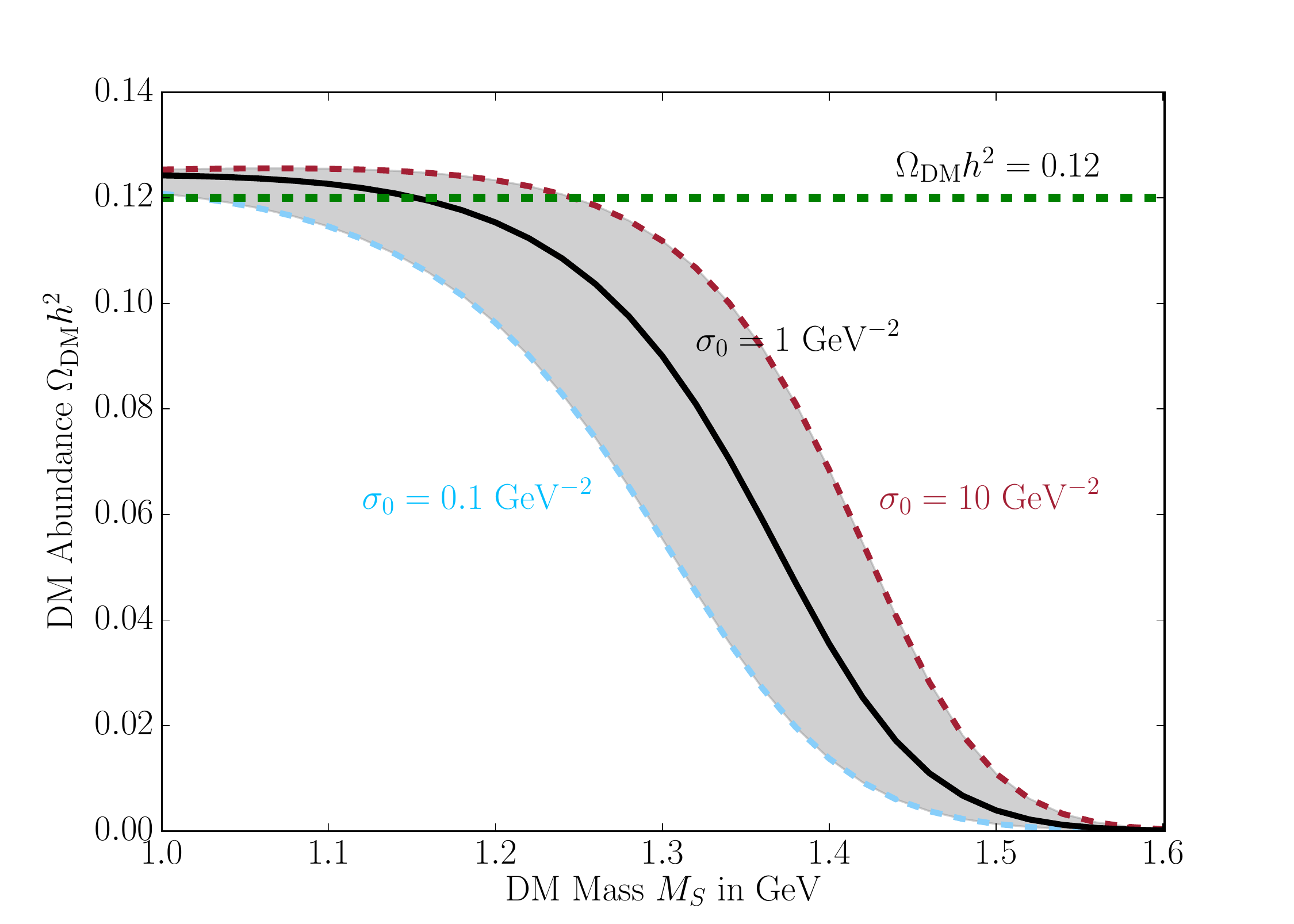}\qquad
\caption{\em \label{fig:YSYb} Thermal hexa-quark abundance within the SM
with the baryon asymmetry in eq.\eq{YBS}.}
\end{center}
\end{figure}

Defining $z = M_\S/T$ and $Y_p = n_p/s$, 
the Boltzmann equation  for the $\S$ abundance is
\beq  \label{eq:nI2}
sHz\frac{dY_\S}{dz} =\gamma_\S^{\rm eqb}
\bigg[\frac{Y^2_B}{Y_B^{\rm eqb2}} -\frac{Y_\S}{Y_\S^{\rm eqb}}  \bigg]\eeq
where the superscript `eqb' denotes thermal equilibrium at fixed baryon asymmetry and 
$Y_B$ is summed over all baryons, but the dibaryon $\S$.
A second equation for $Y_B$ is not needed, given that  baryon number is conserved:
$(Y_B - Y_{\bar B}) +  2 (Y_\S-Y_{\bar \S}) = Y_{B\S}$.
Furthermore, $Y_{\bar\S}$ is negligible, and $Y_{\bar B}$ is negligible around decoupling.
The $\S$ production rate is obtained after summing over all $b b'\leftrightarrow \S X$ processes of eq.\eq{Sprod}.
In the non-relativistic limit the interaction rate gets approximated as
\beq 2\gamma_S^{\rm eqb} \stackrel{T\ll M_\S}{\simeq} \sum_{b,b'}
n_{b}^{\rm eqb}n_{b'}^{\rm eqb} \med{\sigma_{bb'} v_{\rm rel}}^{\rm eqb}
.\eeq
The opposite process is more conveniently written in terms of the $\S$ breaking width defined by
$\gamma_\S^{\rm eqb}(T) = n_S^{\rm eqb}   \Gamma_{\S}^{\rm eqb}$ and given by
\beq
\Gamma_S^{\rm eqb} = \sum_{b,b'} \frac{n_{b}^{\rm eqb}n_{b'}^{\rm eqb}}{2n_S^{\rm eq}} \med{\sigma_{bb'}  v_{\rm rel}}_{\rm eqb}
\eeq
This gets Boltzmann suppressed at $T \circa{<} M_\Lambda - M_p \sim m_s$, when hyperons disappear from the thermal plasma.
Assuming $Y_\S  \circa{<} Y_B$, the Boltzmann equation is approximatively solved by 
$Y_\S \sim Y_\S^{\rm eqb}$ evaluated at the decoupling  epoch where $\Gamma_\S^{\rm eqb} \sim H$,
which corresponds to $T_{\rm dec}\sim  m_s/\ln Y_B $.  
This leads to the estimated final abundance
\beq \frac{Y_\S}{Y_B}\sim Y_B \left(M_{\rm Pl} T_{\rm dec} \sigma_{\S}\right)^{\frac{2M_p-M_\S}{2M_\Lambda-M_\S}}.\eeq
The fact that $\S$ is in thermal equilibrium down to a few tens of MeV means that whatever 
happens at higher temperatures gets washed out.
Notice the unusual dependence on the cross section for $\S$ formation:
increasing it delays the decoupling, increasing the $\S$ abundance.

\smallskip

Fig.\fig{YSYb} shows the numerical result for the relic $\S$ abundance,
computed inserting in the Boltzmann equation a $s$-wave $\sigma_{bb'} v_{\rm rel} = \sigma_0$,
varied around $1/\GeV^2$.
The cosmological DM abundance is reproduced for $M_\S \approx 1.2\GeV$, while
a large $M_\S$ gives a smaller relic abundance.
Bound-state effects at BBN negligibly affect the result, and in particular do not allow to reproduce
the DM abundance with a heavier $M_\S \approx 1.8\GeV$.

\medskip

We conclude this section with some sparse comments.
Possible troubles with bounds from direct detection have been pointed out in~\cite{1711.10971,1802.03025,1802.04764}:
a DM velocity somehow smaller than the expected one can avoid such bounds reducing the kinetic energy available for direct detection.
Using a target made of anti-matter (possibly in the upper atmosphere) would give a sharp annihilation signal, although with small rates.
The magnetic dipole interaction of $\S$ does 
not allow to explain the recent 21 cm anomaly along the lines of~\cite{Barkana:2018lgd}
(an electric dipole would be needed).
The interactions of DM with the baryon/photon fluid may alter the evolution of cosmological perturbations leaving an imprint 
on the matter power spectrum and the CMB.
However, they are not strong enough to produce significant effects.
The $\S$ particle is electrically neutral and has spin zero, such that its
coupling to photons is therefore suppressed by powers of the QCD scale~\cite{1708.08951}. 
So elastic scattering of $\S$ with photons is not cosmologically relevant.

A light $\S$ would affect neutron stars, as they are expected to  contain $\Lambda$ particles,
made stable by the large Fermi surface energy of neutrons.
Then, $\Lambda \Lambda \to \S$ would give a loss of pressure, 
possibly incompatible with the observed existence of neutron stars with mass $2.0 M_{\rm sun}$~\cite{neutronstar}.
However, we cannot exclude $\S$ on this basis, because
production of $\Lambda$ hyperons poses a similar puzzle.
$\S$ as DM could interact with cosmic ray $p$ giving and photon and other signals~\cite{astro-ph/0203240} and
would be geometrically captured in the sun, possibly affecting helioseismology.\footnote{We thank
M. Pospelov for suggesting these ideas.}

In the next section we discuss the main problem which seems to exclude $\S$ as DM.

\begin{figure}
\begin{center}
\includegraphics[width=.25\textwidth]{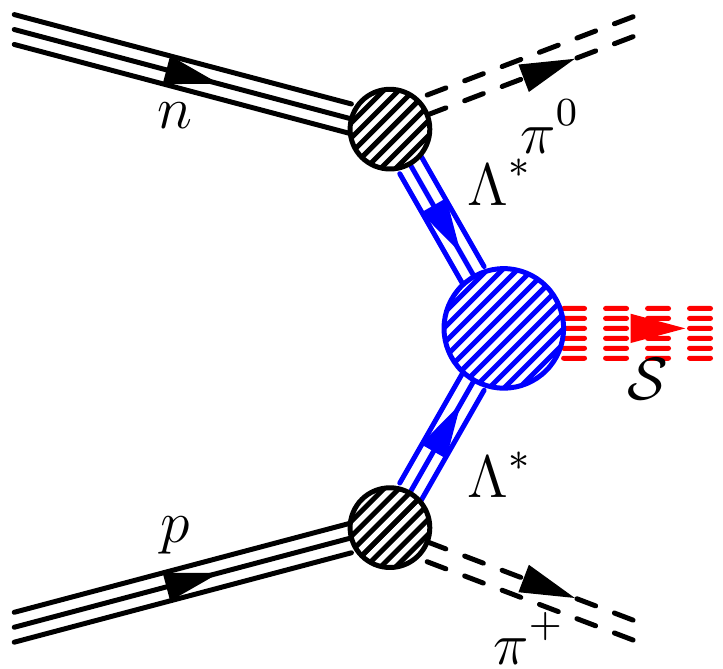}
\end{center}
\caption{\em Sample diagram that dominates nucleon decay into $\S$ inside nuclei. The initial state can also be $nn$ or $pp$.\label{fig:FeynS}}
\end{figure}

\subsection{Super-Kamiokande bound on nuclear stability}\label{SK}

Two nucleons $N=\{n,p\}$ inside a nucleus ${\cal N}$ can make a double weak decay into $\S$,
emitting $\pi$, $\gamma$ or $e$~\cite{Farrar:2003qy}.
This is best probed by Super-Kamiokande (SK), which contains $\sim 8 \times 10^{32}$ Oxygen nuclei.
No dedicated search for $^{16}{\rm O}_8 \to {\cal N'} \, \S \, X$ (where $X$ can be one or two $\pi,\gamma,e$ and ${\cal N'}$ can be $^{14}{\rm O}_8,^{14}{\rm N}_7,^{14}{\rm C}_6$, depending on the charge of $X$) 
has been performed,\footnote{SK searched for di-nucleon decays into pions~\cite{Gustafson:2015qyo} and leptons~\cite{Takhistov:2015fao} and obtained bounds on the lifetime around $\sim 10^{32}$ years.
However these bounds are not directly applicable to $^{16}{\rm O}_8 \to {\cal N'} \, \S \, X$  where the invisible $\S$ takes away most of the energy reducing the energy of the visible pions and charged leptons, in contrast to what is assumed in~\cite{Gustafson:2015qyo,Takhistov:2015fao}.}
but a very conservative limit 
\beq 
\tau({ ^{16}{\rm O}_8 \to {\cal N'} \, \S \, X}) \gtrsim 10^{26}\,{\rm yr}
\eeq
is obtained by requiring the rate of such transitions to
be smaller than the rate of triggered background events in SK,
which is about $10$~Hz~\cite{Fukuda:1998tw}.
A more careful analysis would likely improve this bound by three orders of magnitude~\cite{Farrar:2003qy}.


\begin{figure}
\begin{center}
$$\includegraphics[width=.47\textwidth]{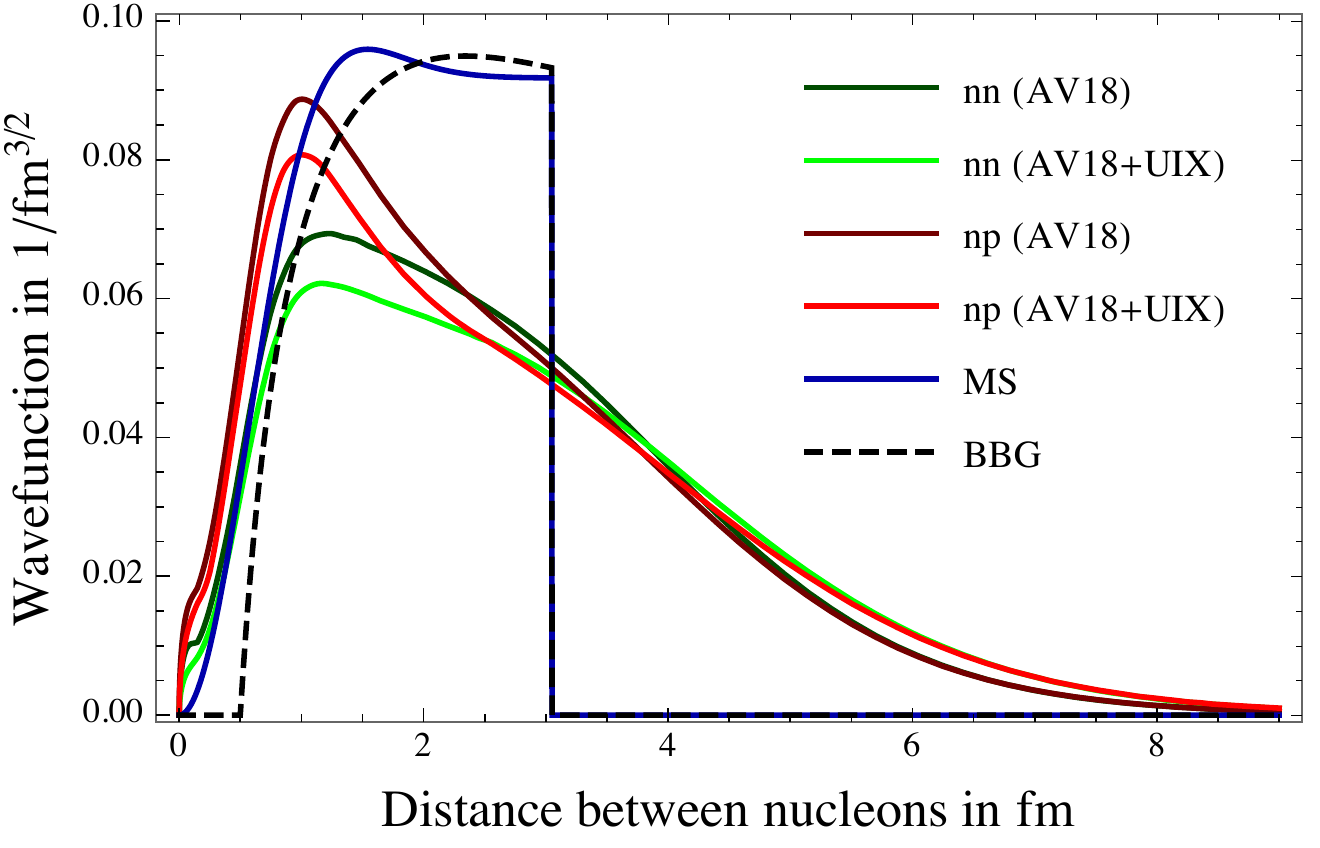}\qquad
\includegraphics[width=.47\textwidth]{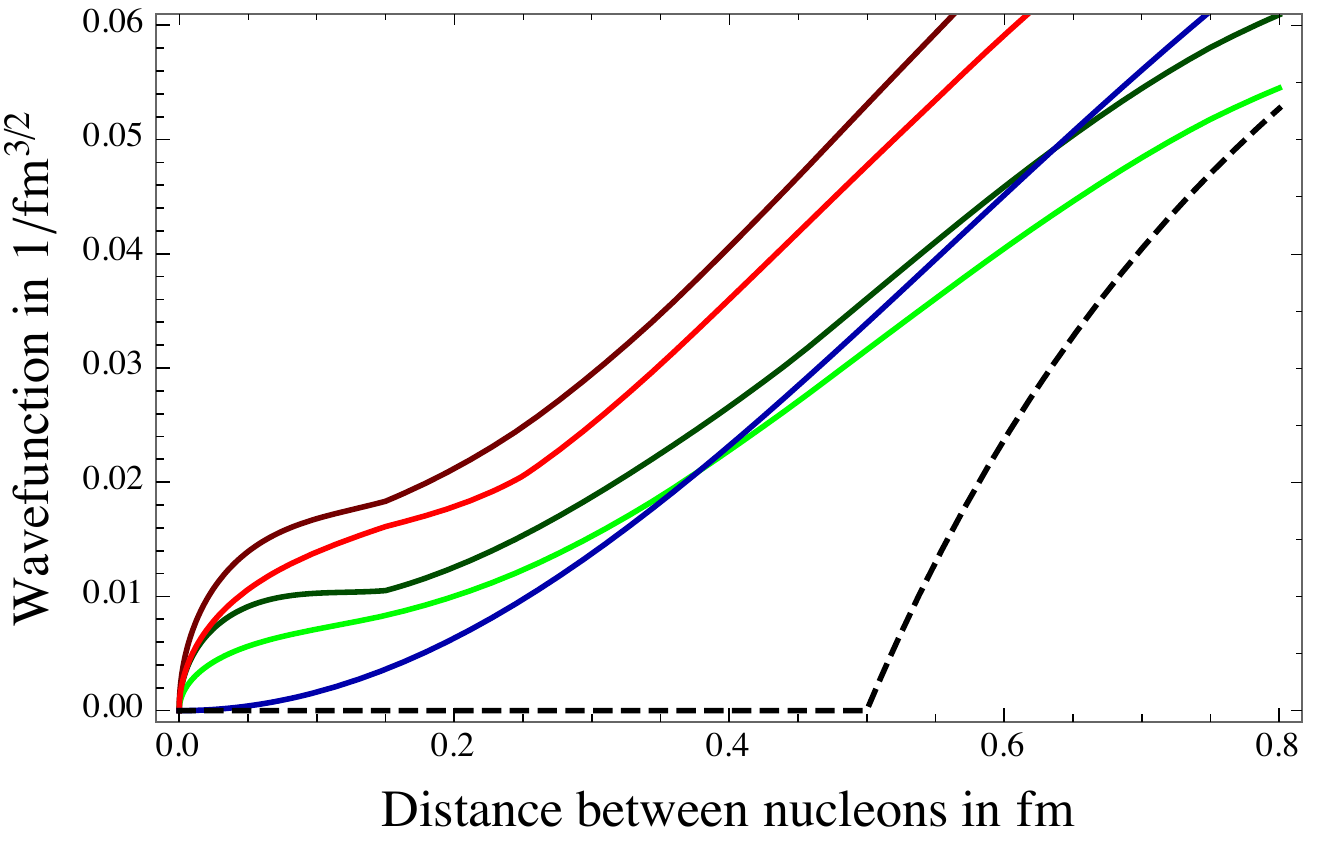}$$
\caption{\em Wave functions $\psi _{\rm nuc}(a)$ as function of
the relative distance $a$ between two nucleons in $^{16}{\rm O}_8$.
The darker (ligther) green curve shows $\psi^{nn}_{\rm nuc}$, obtained from~\cite{Lonardoni:2017egu}, using the AV18 (AV18+UIX) potential, 
the darker (ligther) red curve shows $\psi^{np}_{\rm nuc}$ using the AV18 (AV18+UIX) potential,
the blue curve is the MS wave function used in~\cite{Farrar:2003qy}, 
the dashed black curve is the BBG wave function with hard core radius $r_{\rm core}=0.5\,{\rm fm}$.
{\bf Left:} Wave functions over the entire range.
{\bf Right:} Zoom-in to the region relevant for our calculation.
\label{fig:wf}}
\end{center}
\end{figure}

The amplitude for the formation of $\S$ is reasonably dominated by the sample diagram in fig.\fig{FeynS}:
doubly-weak production of two virtual strange $\Lambda^*$ baryons (e.g.\ through $p\to \pi^+\Lambda^*$ and $n\to \pi^0\Lambda^*$;
at quark level $u\to s\bar d u$ and $d\to s\bar u u$), followed by the strong process
$\Lambda^*\Lambda^*\to \S$:
\beq 
{\cal M}_{NN \rightarrow \S X} \approx
{\cal M}_{NN \rightarrow \Lambda^* \Lambda^* X} \times {\cal M}_{\Lambda^*\Lambda^* \rightarrow \S} .
\eeq
The predicted life-time is then obtained as~\cite{Farrar:2003qy}\footnote{A numerical factor of 1440 due to spin and flavor effects has already been factored out from $|{\cal M}|^2_{\Lambda^*\Lambda^* \rightarrow \S}$ here and in the following.
Note also that the threshold $2 M_N - M_\pi = 1.74$ GeV neglects the small difference in binding energy between ${\cal N}$ and ${\cal N'}$.}
\be
\tau({\cal N}\to {\cal N'} \, \S \, X) \simeq \frac{\textrm{yr}}{|{\cal M}|^2_{\Lambda^*\Lambda^* \rightarrow \S}}
\times
\left\{\begin{array}{ll}
3 & \hbox{if $M_\S \lesssim 1.74\GeV$}\\
10^5 & \hbox{if $1.74\GeV\lesssim M_\S \lesssim1.85\GeV $}
\end{array}\right. 
\ee
where the smaller value holds if $\S$ is so light that the decay can proceed through real $\pi$ or $\pi\pi$ emission, while the longer life-time if obtained if instead only lighter $e^+\nu$ or $\gamma$ can be emitted.

\begin{figure}[t]
\begin{center}
\includegraphics[width=.69\textwidth]{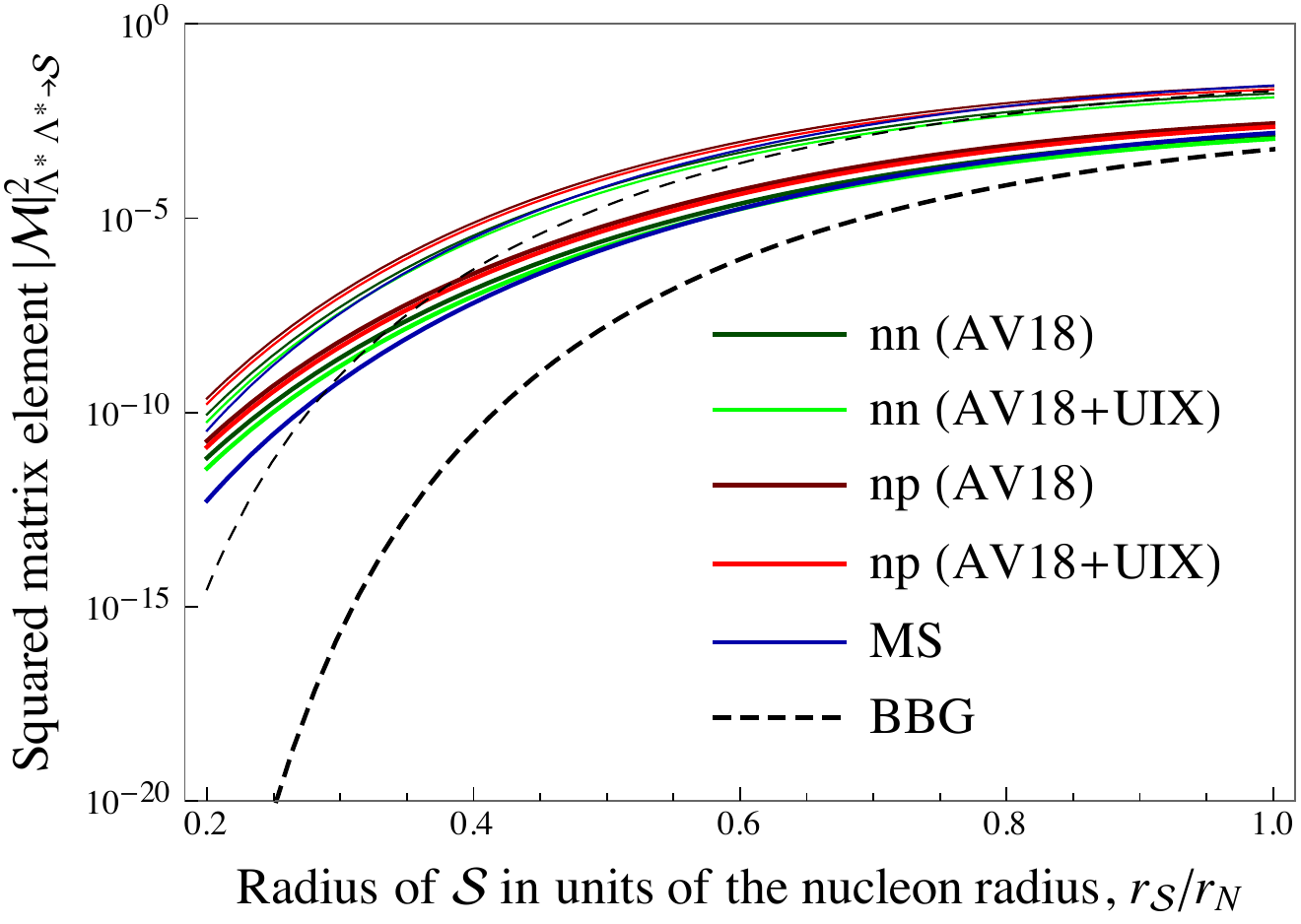}
\caption{\label{fig:MLLtoS} 
\em The dimension-less squared matrix element for nuclear decay into $\S$,
$|{\cal M}|^2_{\Lambda^*\Lambda^* \rightarrow \S}$, as a function of the $\S$ radius $r_\S$ in units of 
the nucleon radius $r_N$, using different nuclear wave functions.
The color coding, defined in fig.~\ref{fig:wf}, refers to the nuclear wave functions used.
The thinner (thicker) curves assume $r_N = 0.87\,{\rm fm}$ ($r_N= 0.49\,{\rm fm}$).
The Super Kamiokande bound would be evaded for 
$|{\cal M}|^2_{\Lambda^*\Lambda^*\to\S}\circa{<} 10^{-20}~(10^{-25})$ for $M_\S >1.74\GeV$
(for $M_\S <1.74\GeV$).}
\end{center}
\end{figure}

The key factor is the dimension-less matrix element ${\cal M}_{\Lambda^*\Lambda^* \rightarrow \S}$ for the transition $\Lambda^* \Lambda^* \rightarrow \S$ inside a nucleus, that we now discuss.
Following~\cite{Farrar:2003qy}, we assume that the initial state wave function can be factorized into wave functions of the two $\Lambda^*$ baryons and a relative wave function $\psi _{\rm nuc}(\vec a)$ for the separation $\vec a$ between the center of mass of the $\Lambda^*$'s.
The matrix element is given by the wave-function overlap
\bal
{\cal M}_{\Lambda^*\Lambda^* \rightarrow \S}
=
 \int \!  \psi^*_\S(\vec\rho{}\,^a,\vec \lambda^a,\vec\rho{}\,^b, \vec \lambda^b,\vec a) \ \psi  _{\Lambda^*}(\vec\rho{}\,^a,\vec \lambda^a) \psi _{\Lambda^*}(\vec\rho{}\,^b,\vec\lambda^b) \psi_{\rm nuc}( \vec a) \,
 d^3\! a \,d^3\! \rho^a d^3\! \rho^b \, d^3\! \lambda^a  d^3\! \lambda^b \,.
\eal
Here, $\vec {\rho}{}\,^{a,b}, \vec {\lambda}^{a,b}$ are center-of-mass coordinates which
parametrise the relative positions of the quarks within each $\Lambda^*$.
Using the Isgur-Karl (IK) model~\cite{Isgur:1978wd} the wave functions for the quarks inside the $\Lambda^*$ and inside the $\S$ are approximated by
\bal
\psi_{\Lambda^*} (\vec {\rho}, \vec {\lambda})
&=\left( \frac
{1}{r_N\sqrt{\pi}} \right) ^3 \exp\left[ -\frac
{\vec \rho{}\,^2 + \vec \lambda^2}{2r_N^2} \right],
\\
\psi_{\S} (\vec {\rho}{}\,^{a}, \vec {\lambda}^{a},\vec {\rho}{}\,^{b}, \vec {\lambda}^{b},\vec {a})
&=\left( \frac{3}{2}\right) ^{3/4}
\left( \frac{1}{r_\S\sqrt{\pi}} \right)^{15/2} \exp\left[-\frac {\vec {\rho}\,^{a2} + \vec
{\lambda}^{a2}+\vec {\rho}\,^{b2} + \vec {\lambda }^{b2} +\frac
{3}{2} \vec {a}^2}{2r_\S^2}\right]  \,,
\eal
where $r_N$  
and $r_\S$ are the radii of the nucleons respectively of $\S$.\footnote{One should be aware that the IK model has serious shortcomings. One issue is that it is a non-relativistic model --- an assumption which is problematic in particular for small $\S$.
Another problem is that the value of $r_N$ that gives a good fit to the lowest lying ${1\over 2}^+$ and ${3\over 2}^+$ baryons --- $r_N=0.49$~fm --- is smaller than the charge radius of the proton: $r_N=0.87$~fm.
Therefore we consider both $r_N=0.49$~fm and $r_N=0.87$~fm, as done in~\cite{Farrar:2003qy}.} 
Performing all integrals except the final integral over $a\equiv |\vec a|$ gives
\bal
|{\cal M}|_{\Lambda^*\Lambda^* \rightarrow \S}=
\frac {1}{2}  \left( \frac{3}{2}\right)^{3/4}
\left(\frac {2 r_N r_\S}{r_N^2+r_\S^2}\right )^6 \left( \frac{1}{r_\S\sqrt{\pi}} \right)^{3/2}
\int d a \ 4 \pi a^2
e ^{-3a^2/4r_\S^2} \psi _{\rm nuc}(a) \,.
\eal
As shown in fig.~\ref{fig:MLLtoS} below (and as discussed in~\cite{Farrar:2003qy}), if $r_\S$ is not much smaller than $r_N$, the overlap integral is not very much suppressed and $\tau({ ^{16}{\rm O}_8 \to {\cal N'} \, \S \, X})$ is tens of orders of magnitude below the experimental limit, and is clearly excluded.
This conclusion is independent of the form of $\psi _{\rm nuc}$.

However if $r_\S$ were a few times smaller than $r_N$ --- a possibility which seems unlikely due to diquark repulsions (see e.g.~\cite{Selem:2006nd}) but cannot firmly be excluded --- then $\tau({ ^{16}{\rm O}_8 \to {\cal N'} \, \S \, X})$ is extremely sensitive to the probability of the overlap of two nucleons inside the oxygen core at very small distances (less than, say, 0.5~fm).
The wave function of nucleon pairs $\psi _{\rm nuc}$ at such small distances has not been probed experimentally. 
In fact, at such small distances nucleons are not the appropriate degrees of freedom.\footnote{Data indicate that about 20\% of the nucleons form $pn$ pairs so close (about 1 fm) that the local density reaches the nucleon density (about 2.5 times larger than the nuclear density) and thus that  the quark structure of nucleons starts becoming relevant already at $a \sim$ 1 fm~\cite{1611.09748}.}
Thus, for a very small $\S$ one can only make an educated guess of $\tau({ ^{16}{\rm O}_8 \to {\cal N'} \, \S \, X})$, since the form of $\psi _{\rm nuc}$ is uncertain.
Nevertheless, we will show in the following that for a reasonable form of $\psi _{\rm nuc}$ a stable $\S$ is excluded even if it were very small.

Numerical computations of the ground-state wave-functions of nuclei, including $^{16}{\rm O}_8$ have been performed e.g. in~\cite{Lonardoni:2017egu}.
The quantity that determines $\psi _{\rm nuc}$ is the two-nucleon point density $\rho_{NN}$, defined in eq.~(58) of~\cite{Lonardoni:2017egu}.
We obtain $\rho_{NN}(a)$ by interpolating the data given in~\cite{Lonardoni:2017egu} and adding the constraint $\rho_{NN}(0)=0$, which is
a conservative assumption for our purposes since 
$\rho_{NN}(0)\neq0$ would lead to a larger matrix element.
There are 28 neutron-neutron pairs and 64 proton-neutron pairs in $^{16}{\rm O}_8$ so one has $\int da \, 4 \pi a^2 \rho_{nn}(a)=28$ and $\int da \, 4 \pi a^2 \rho_{pn}(a)=64$.
We therefore define the wavefunctions
\be
\psi^{nn} _{\rm nuc}(a)=\sqrt{\rho_{nn}(a)/28} \,, \quad \psi^{pn} _{\rm nuc}(a)=\sqrt{\rho_{pn}(a)/64} \,.
\ee
These wave functions are plotted in fig.~\ref{fig:wf}, together with 
the Miller-Spencer (MS) and the Brueckner-Bethe-Goldstone (BBG) wave function  used in~\cite{Farrar:2003qy}.
The BBG wave functions assume 
a hard repulsive core between nucleons such that $\psi_{\rm nuc}$ vanishes at $a<r_{\rm core}$.
We take $r_{\rm core} =0.5\,{\rm fm}$ for illustration.
This is not realistic but allows to see what kind of nuclear wave function would sufficiently suppress the rate of $\S$-formation in nuclei, if $\S$ is small enough.
The resulting $|{\cal M}|^2_{\Lambda^*\Lambda^* \rightarrow \S}$ is plotted in fig.~\ref{fig:MLLtoS}, again compared to that obtained using the Miller-Spencer and BBG wave functions.


The resulting matrix elements from the MS wave function qualitatively agree to what is obtained using the wave functions extracted from~\cite{Lonardoni:2017egu}.
By contrast, the matrix element using the BBG wave function with hard core radius $r_{\rm core}=0.5\,{\rm fm}$ is very much suppressed, especially if $\S$ is small. 
The reason is that, according to the assumption of a hard core repulsive potential, 
the nucleons can't get close enough to form the small state $\S$.
Since we do not consider a $\psi_{\rm nuc}(a)$ which vanishes for $a \lesssim 0.5$~fm realistic, we conclude that a stable $\S$ is excluded.
%

%

Weaker bounds on $\S$ production are obtained considering baryons containing $\Lambda$'s.

 \begin{figure}[t]
\begin{center}
$$\includegraphics[width=.46\textwidth]{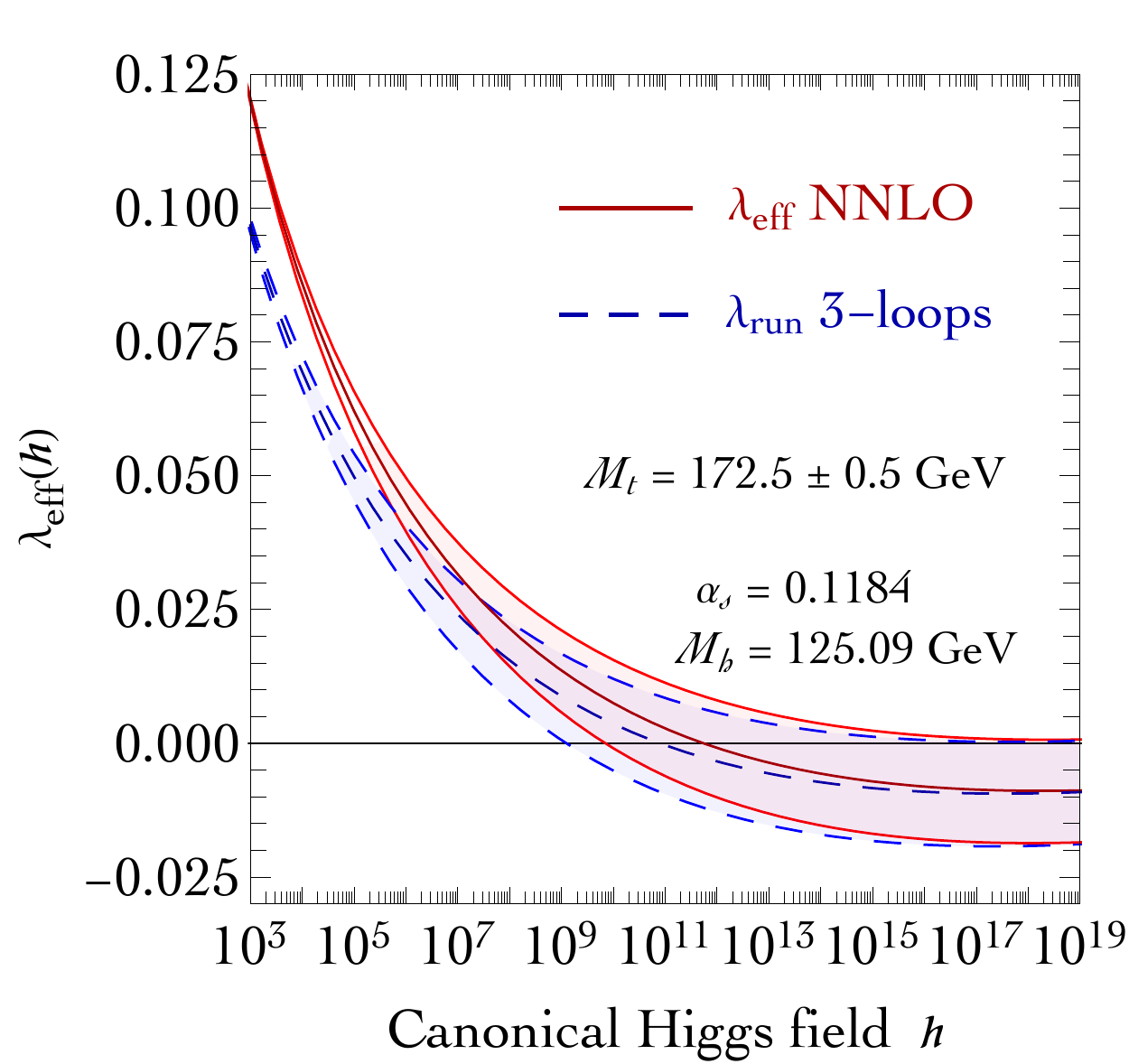}\qquad
\includegraphics[width=.44\textwidth]{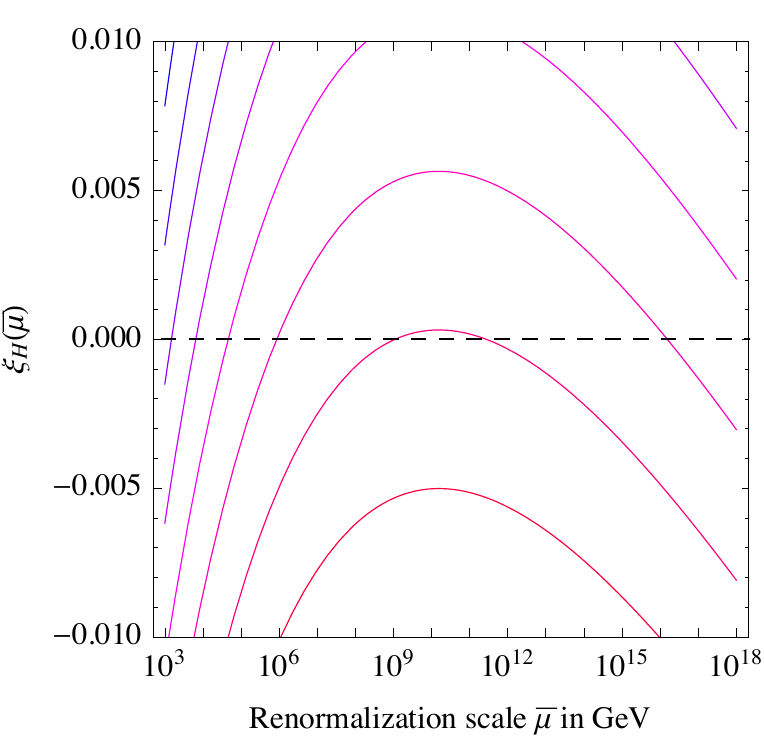}$$
\caption{\em \label{fig:Running}  
{\bf Left}: RGE running of $\lambda_{\rm eff}$ (solid red lines) obtained by changing the top mass in its $3\sigma$ interval defined by $M_t = \Mtexp$.
 For comparison, we show the running of $\lambda$ at 3-loops (dashed blue lines) without including the Coleman-Weinberg corrections.
{\bf Right}: RGE running of a small $\xi_H$.
}
\end{center}
\end{figure}

\section{DM as black holes triggered by Higgs fluctuations}\label{DMBH}
We here present the technical computations relative to the mechanism anticipated in the Introduction.
The SM potential is summarized in section~\ref{Veff}.
In section~\ref{sec:BGdyn} we outline the mechanism that generates black holes.
Section~\ref{sec:BeforeInf} studies the generation of Higgs inhomogeneities.
Post-inflationary dynamics is studied in section~\ref{postinf}.
Formation of black holes is considered in section~\ref{BH}.
The viability of a critical assumption is discussed in section~\ref{Homo}.

\subsection{The Higgs effective potential}\label{Veff}
The effective potential of the canonically normalised
Higgs field during inflation with Hubble constant $H_0$
is 
\begin{equation}\label{eq:EffPotInf}
V_{\rm eff}(h) \approx \frac{\lambda_{\rm eff}(h)}{4}h^4 - 6\xi_H H_0^2 h^2 + V_0~,
\end{equation}
at $h \gg 174\GeV$.
Here $\lambda_{\rm eff}$ is the effective quartic coupling computed including quantum corrections. 
The second mass term in $V_{\rm eff}(h)$ can be generated by various different sources~\cite{Espinosa:2017sgp}.
We consider the minimal source: a Higgs coupling to gravity, $\mathscr{L}_{\xi} = -\xi_HRh^2/2$, with Ricci scalar 
$R = -12H_0^2$ during inflation. 
Finally, during inflation the effective potential in eq.~(\ref{eq:EffPotInf}) is augmented by the 
vacuum energy associated to the inflaton sector, $V_0 = 3\bar{M}_{\rm Pl}^2 H_0^2$, where 
 $\bar{M}_{\rm Pl} \simeq 2.435\times 10^{18}$ GeV is the reduced Planck mass.
 

We implement the RG-improvement of the  effective potential at NNLO precision:
running  the SM parameters at 3-loops  and
including $2$-loop quantum corrections to the effective potential.
 We consider fixed values of $\alpha_3(M_Z) = 0.1184$ and $M_h = 125.09$ GeV, 
and we vary the main uncertain parameter, the top mass, in the interval 
$M_t = \Mtexp$~\cite{Mtexp}.
In fig.~\ref{fig:Running}a we show the resulting $\lambda_{\rm eff}(h)$ as function of $h$.  

\medskip

The non-minimal coupling to gravity $\xi_H$ receives  SM quantum corrections encoded in its RGE,
which induce $\xi_H \neq 0$ even starting from $\xi_H = 0$ at some energy scale. 
The RGE running of small values of $\xi_H(\bar{\mu})$ is shown in  fig.~\ref{fig:Running}b.
As mentioned before, a non-zero $\xi_H$ 
can be considered as a proxy for an effective mass term during inflation.
The latter, for instance, can be generated by a quartic interaction $\lambda_{h\phi}|H|^2\phi^2$ between the Higgs and the inflaton field or by the inflaton decay into SM particles during inflation.
For this reasons, it makes  sense to include $\xi_H$ as a free parameter in the analysis of the Higgs dynamics during inflation, at most with the theoretical bias that its size could be loop-suppressed.

\medskip

\subsubsection*{Analytic approximation}
\begin{figure}[t]
\begin{center}
$$\includegraphics[width=.434\textwidth]{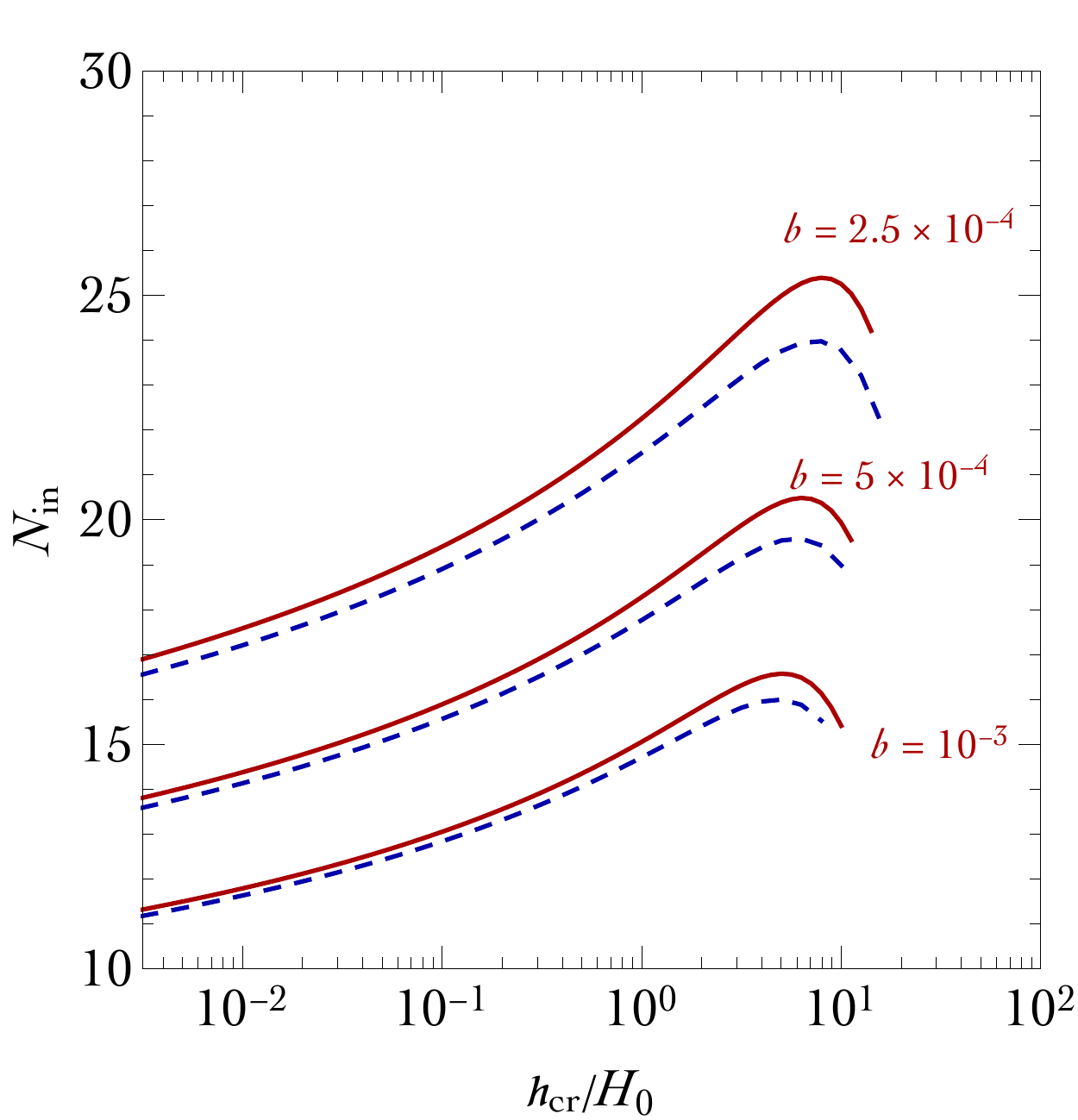}
\qquad\includegraphics[width=.45\textwidth]{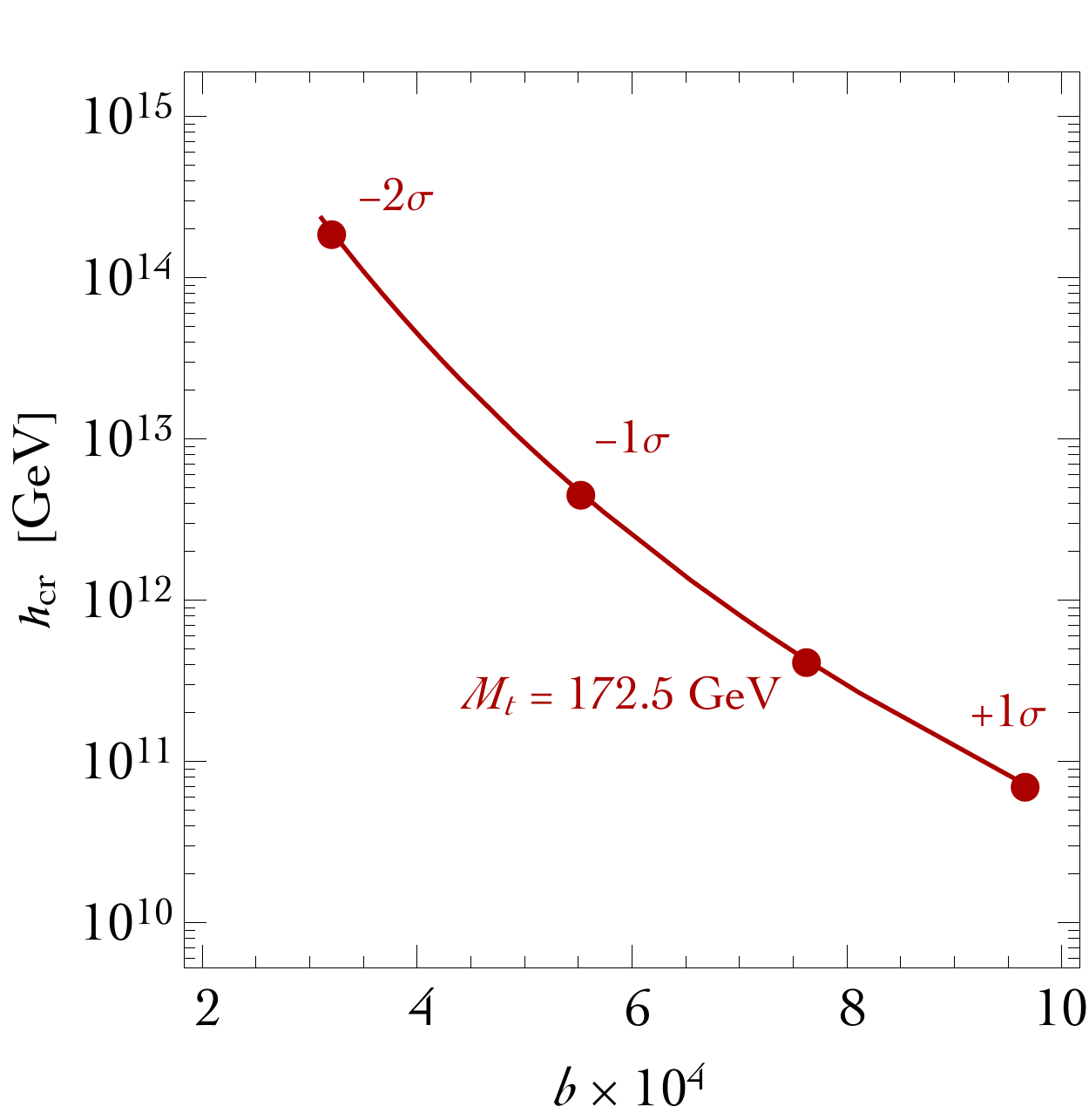}$$
\caption{\em \label{fig:Corr} 
{\bf Left}: 
Number of $e$-folds $N_{\rm in}$ at the beginning of the Higgs fall that gives the maximal $h_{\rm end}$ rescued by
the reheating temperature.  This is computed as function of the Hubble constant during inflation, for 3 different values
of the uncertain parameter $b$ that approximates the Higgs potential.  
Continuous (dashed) curves correspond to $\xi_H=0$ ($-10^{-3}$).
{\bf Right}:  SM values of $b$ and
of the position $h_{\rm cr}$ of the top of the SM potential as function of the top mass.  }
\end{center}
\end{figure}

We will show precise numerical results for the SM case.
However the discussion is clarified by introducing a simple approximation that
encodes the main features of the SM effective potential in eq.~(\ref{eq:EffPotInf}):
\begin{equation}\label{eq:Potential}
V_{\rm eff}(h) \approx -b\log\left(
\frac{h^2}{h_{\rm cr}^2\sqrt{e}}
\right)\frac{h^4}{4} - 6\xi_{H}H_0^2h^2~, 
\end{equation}
where $h_{\rm cr}$ is the position of the maximum of the potential with no extra mass term, $\xi_H = 0$.
The parameters $b$ and $h_{\rm cr}$ depend on the low-energy SM parameters such as the top mass:
they can be computed by matching the numerical value of the Higgs effective potential at the gauge-invariant position of the maximum, $V_{\rm eff}(h_{\rm cr}) = bh_{\rm cr}^4/8$.
The result is shown in the right panel of fig.~\ref{fig:Corr}. 

Results will be better understood when presented in terms of the dimensionless parameters
$b$, $\xi_H$, $h_{\rm cr}/H_0$ and $T/H_0$, where $T$ is the temperature,
as they directly control the dynamics that we are going to study.
The parameter $b$ controls the flatness of the potential beyond the potential barrier at $h_{\rm cr}$,  with smaller $b$ corresponding to a flatter potential.
The non-minimal coupling $\xi_H$ controls the effective Higgs mass during inflation.
Finally $\bar{M}_{\rm Pl}/H_0$ will set the reheating temperature in eq.~(\ref{eq:ReheatingT}) 
and thus the position and size of the thermal barrier.

\smallskip

The position of the potential barrier --- defined by the field value where the effective potential
 has its maximum --- strongly depends on the value of the top mass, on the non-minimal coupling to gravity, and, after inflation, on
 the temperature of the thermal bath which provides and extra mass term.
For $\xi_{H} \neq 0$, the maximum of the Higgs potential gets shifted from $h_{\rm cr}$ to 
\begin{equation}
h_{\rm max} = H_0 \left[
-\frac{b}{12\xi_H}\mathcal{W}\left(
\frac{-12\xi_H H_0^2}{b h_{\rm cr}^2}
\right)
\right]^{-1/2}~,
\end{equation}
where $\mathcal{W}(z)$ is the product-log function defined by
 $z=\mathcal{W}e^\mathcal{W}$. 
The condition
 \begin{equation}
 -12\xi_H H_0^2 > -\frac{bh_{\rm cr}^2}{e}~,
 \end{equation}
must be satisfied otherwise the effective mass is too negative and it erases the potential barrier, thus leading to a classical instability.

\subsubsection*{The thermal potential}
After the end of inflation, the Higgs effective potential receives large 
thermal corrections from the SM bath at generic temperature $T$.
The initial temperature of the thermal bath is fixed by the dynamics of reheating after inflation.
We  assume instantaneous reheating, as this is most efficient for rescuing the falling Higgs field.
The reheating temperature is then given by 
\begin{equation}\label{eq:ReheatingT}
T_{\rm RH} = \left(
\frac{45}{4\pi^3 g_*}
\right)^{1/4}M_{\rm Pl}^{1/2}H_0^{1/2}~,
\end{equation}
where $g_* = 106.75$ is the number of SM degrees of freedom. 
After reheating the Universe becomes radiation-dominated, the Ricci scalar vanishes, and so the contribution to the effective potential from the non-minimal Higgs coupling to gravity.

The effective Higgs potential at  finite temperature is obtained
adding an extra thermal contribution $V_{\rm T}$
which can be approximated as an effective thermal mass for the Higgs field, $M_{\rm T}^2 \simeq 0.12\,T^2$
(see e.g.~\cite{1505.04825})
\begin{equation}\label{eq:ThermalPot}
V_{\rm eff}^{\rm T}(h) \approx -b\log\left(
\frac{h^2}{h_{\rm cr}^2\sqrt{e}}
\right)\frac{h^4}{4} + V_{\rm T}(h)~,
\qquad
V_{\rm T}(h) \approx \frac{1}{2}M_{\rm T}^2 h^2 e^{-h/2\pi T}~.
\end{equation}
At $h \circa{<} T$ we can neglect the exponential suppression in the thermal mass, and
the maximum of the effective potential in eq.~(\ref{eq:ThermalPot}) is given by
\begin{equation}\label{eq:ThermalMax}
h_{\rm max}^{\rm T} = M_{\rm T}\left[
 b\,\mathcal{W}\left(
\frac{M_{\rm T}^2}{bh_{\rm cr}^2}
\right)
\right]^{-1/2}.
\end{equation}

\subsection{Outline of the mechanism}\label{sec:BGdyn}
During inflation, the Higgs field is subject to quantum fluctuations. 
Depending on the value of $H_0$, 
these quantum fluctuations could lead the Higgs
 beyond the barrier, and make it roll towards Planckian values. 
If $T_{\rm RH}$ 
is high enough and $h$ is not too far, thermal corrections can
 ``rescue'' the Higgs, bringing it back to the origin~\cite{1505.04825}. 
The mechanism relies on a tuning such that the following situation occurs~\cite{Espinosa:2017sgp}:
\begin{enumerate}[{\it i)}]
  \item At $N_{\rm in}\sim 20$ $e$-folds before the end of inflation,
the Higgs background value $h$ is brought by quantum fluctuation to some $h_{\rm in}\neq 0$. 
This configuration must be spatially homogeneous on an inflating local
 patch large enough to encompass our observable Universe today.
We consider the de Sitter metric in flat slicing coordinates,  $ds^2 = -dt^2 + a^2(t)d\vec{x}^2$.
 We will discuss later how precisely this assumption must be satisfied, and its plausibility.

  \item When the classical evolution prevails over the quantum corrections, the Higgs field,
  starting from the initial position $h_{\rm in}$, 
   begins to slow roll down the negative potential. 
   This condition reads 
\begin{equation}\label{eq:ClassicalvsQuantum}
\underbrace{\left|\frac{V_{\rm eff}^{\prime}(h_{\rm in})}{3H_0^2}\right|}_{\rm classical} 
> \underbrace{\frac{cH_0}{2\pi}}_{\rm quantum}~.
\end{equation}
where $c$ is a constant of order 1, fixed to $c=1$ in~\cite{Espinosa:2017sgp}.
We will explore what happens choosing $c=0.9$ or $c=1.1$.
   From this starting point $t_{\rm in}$ on, 
   the classical evolution of the background Higgs value is
  described by 
\begin{equation}\label{eq:ClassicalMotion}
 \ddot{h}_{\rm cl} + 3H_0\dot{h}_{\rm cl} + V^{\prime}_{\rm eff}(h_{\rm cl}) = 0
 \end{equation}
where  the subscript $_{\rm cl}$ indicates that this is a classical motion. 
Dots indicate derivatives with respect to time $t$.

 \item At the end of inflation, $N_{\rm end} = 0$, the Higgs is rescued by thermal effects.
This happens if the value of the Higgs field $h_{\rm end}$ at the end of inflation 
is smaller than   the position of the thermal potential barrier at reheating, $h_{\rm max}^{T_{\rm RH}}$.
A significant amount of PBH  arises only if this condition is barely satisfied in all Universe.
This is why the homogeneity assumption in $i)$ is needed.
 \end{enumerate}
 To compute condition $iii)$ we fix the initial value of the classical motion  $h_{\rm in}$ such that
eq.~(\ref{eq:ClassicalvsQuantum}) is satisfied with $c = 1$;
next, we maximise the $h_{\rm end}$ obtained solving eq.~(\ref{eq:ClassicalMotion})
by tuning the amount of inflation where the fall happens, as parameterized by $N_{\rm in}$.
 The left panel of fig.~\ref{fig:Corr} shows the 
 initial value $N_{\rm in}$ obtained following this procedure as a function of $h_{\rm cr}$ in units of $H_0$. 
 Smaller values of $b$ (i.e. smaller values of $M_t$) imply a flattening of the potential, and the classical dynamics during inflation is slower. The right side of the curves is limited by the classicality condition 
 in eq.~(\ref{eq:ClassicalvsQuantum}).
A $\xi_H <0$ shifts the position of the potential barrier towards the limiting value $h_{\rm max}^{\rm T}$ in eq.~(\ref{eq:ThermalMax}) --- which does not depend on $\xi_H$ --- above which the rescue mechanism due to thermal effects is no-longer effective: its net effect is to reduce the number of $e$-folds during which classical motion can happen (for fixed $b$).

\medskip 
 
We anticipate here the feature of PBH formation which implies the
restriction on the parameter space mentioned at point $i)$: Higgs fall must start at least $N_{\rm in}\sim 20$
$e$-folds before the end of inflation. 
The collapse of the mass inside the horizon $N$ $e$-folds before inflation end forms a PBH with mass (see also section~\ref{BH})
\begin{equation}\label{eq:PBHmass}   M_{\rm PBH} \approx \frac{\bar{M}_{\rm Pl}^2}{H_0}e^{2N}~.
   \end{equation}
PBH  must be heavy enough to avoid Hawking evaporation. 
The lifetime of a PBH with mass $M_{\rm PBH}$ due to Hawking radiation at Bekenstein-Hawking temperature 
$
T_{\rm BH} = \sfrac{1}{(8\pi G_N M_{\rm PBH})} 
$
 is 
 \begin{equation}
 \Gamma^{-1}_{\rm PBH} \approx 4\times 10^{11}\left[
 \frac{\mathcal{F}(M_{\rm PBH})}{15.35}
 \right]^{-1}\left(
 \frac{M_{\rm PBH}}{10^{13}\,{\rm g}}
 \right)^{3}\,{\rm s}~,
 \end{equation}
where $\mathcal{F} \to 1$ at $M_{\rm PBH} > 10^{17}$ g. 
BH heavier than $M_{\rm PBH} > 10^{15}$ g are cosmologically stable, 
and BH heavier than $M_{\rm PBH} > 10^{16.5}$ g are allowed by bounds on Hawking radiation
as a (significant fraction of) DM.
Since $N < N_{\rm in}$,  imposing   $M_{\rm PBH} > 10^{16.5}$ g 
implies a conservative lower limit on $N_{\rm in}$:
 \begin{equation}
 N_{\rm in} > \frac{1}{2}\ln\left[  7.2\times 10^{21} \frac{H_0}{\bar{M}_{\rm Pl}} \right] =  
 18.3\,\,\,\,\,\hbox{ for $H_0  = 10^{-6}\bar{M}_{\rm Pl}$}. 
 \end{equation}

\subsection{Higgs fluctuations during inflation}\label{sec:BeforeInf}
We now consider the evolution of Higgs perturbations during inflation.
Expanding $h$ in Fourier space with comoving wavenumber $k$,\footnote{The comoving wavenumber $k = |\vec{k}|$ 
is time independent, and it is related to the physical momentum 
via $k_{\rm phys} = k/a(t)$, which decreases as the space expands.} the equation
 for the mode $\delta h_k$ takes the form
 \begin{equation}
 \ddot{\delta h}_k + 3H_0\dot{\delta h}_k + \frac{k^2}{a^2}\delta h_k + V_{\rm eff}^{\prime\prime}(h_{\rm cl})\delta h = 0~,
 \end{equation}
where we neglected metric fluctuations. 
In terms of the number of $e$-folds $N$ and of the Mukhanov-Sasaki variable $u_k \equiv a\,\delta h_k$  
it becomes
\begin{equation}\label{eq:MS}
\frac{d^2 u_k}{dN^2} + \frac{du_k}{dN} + \left(
\frac{k^2}{a^2 H_0^2} - 2
\right)u_k + \frac{V_{\rm eff}^{\prime\prime}(h_{\rm cl})}{H_0^2}u_k = 0~.
\end{equation}
It is convenient to consider the evolution of the perturbation 
 making reference to a specific moment  before the end of inflation: at the initial value $N_{\rm in}$ defined in section~\ref{sec:BGdyn}. 
We recall that in our convention $N_{\rm end} = 0$ at the end of inflation.
Eq.~(\ref{eq:MS}) becomes 
\begin{equation}\label{eq:MS2}
\frac{d^2 u_k}{dN^2} - \frac{du_k}{dN} + \left[
\left(
\frac{k}{a_{\rm in} H_0}
\right)^2 e^{2(N - N_{\rm in})} - 2
\right]u_k + \frac{V_{\rm eff}^{\prime\prime}(h_{\rm cl})}{H_0^2}u_k = 0~.
\end{equation}
In this form, the Mukhanov-Sasaki equation is particularly illustrative. 
Consider the evolution of the perturbation for a mode of interest $k$ that we fix compared to the reference 
 value $a_{\rm in}H_0$ at $t=t_{\rm in}$. In particular, we consider the case of a mode $k$ that is 
 sub-horizon at the beginning of the classical evolution, that is $k \gg a_{\rm in}H_0$.
 From eq.~(\ref{eq:MS2}), we see that in the subsequent evolution with $N < N_{\rm in}$ the exponential suppression 
 will turn the mode from 
sub-horizon to super-horizon. 

\begin{figure}[t]
\begin{center}
  $$\includegraphics[width=.45\textwidth]{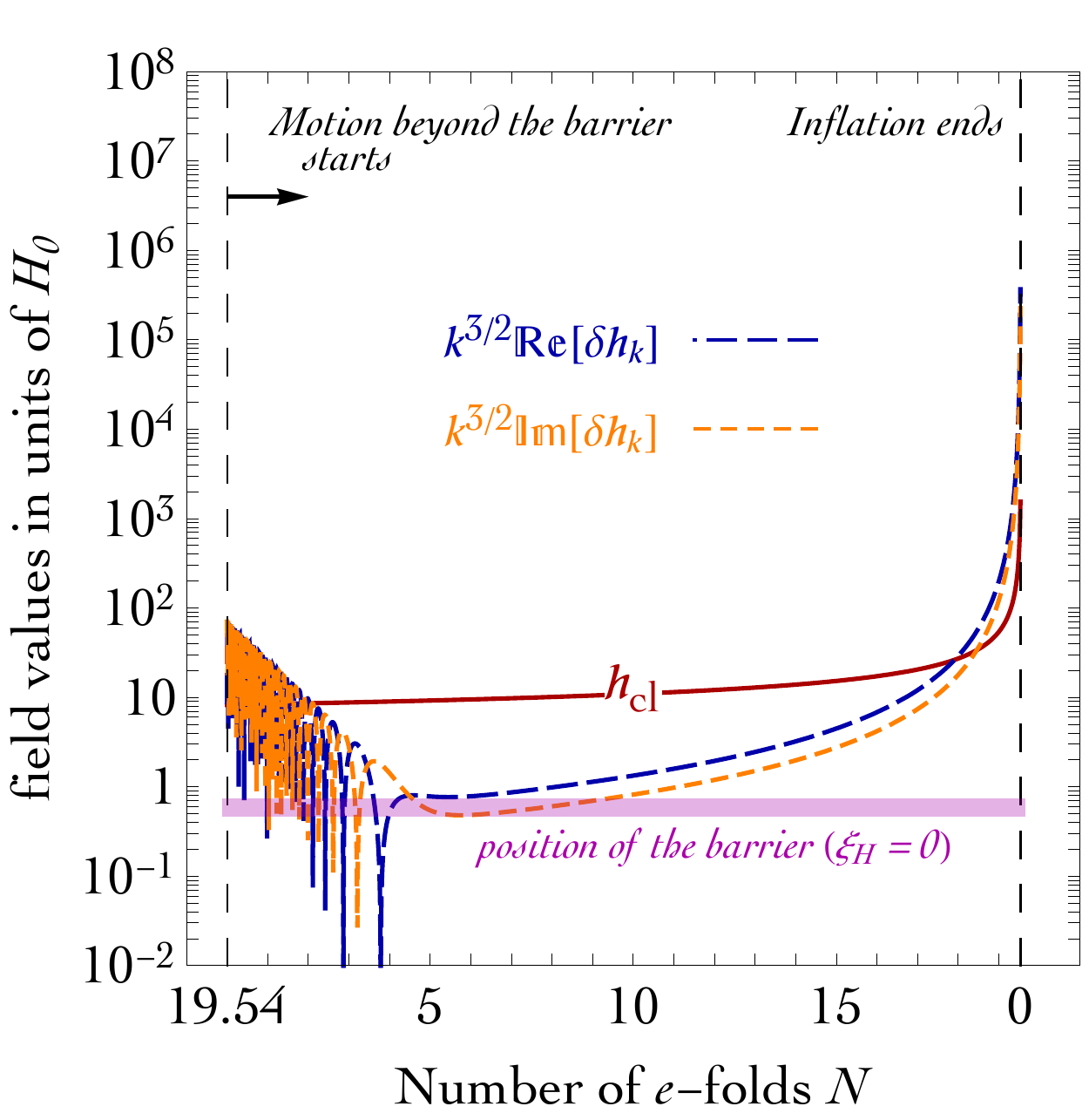}\qquad
  \includegraphics[width=.45\textwidth]{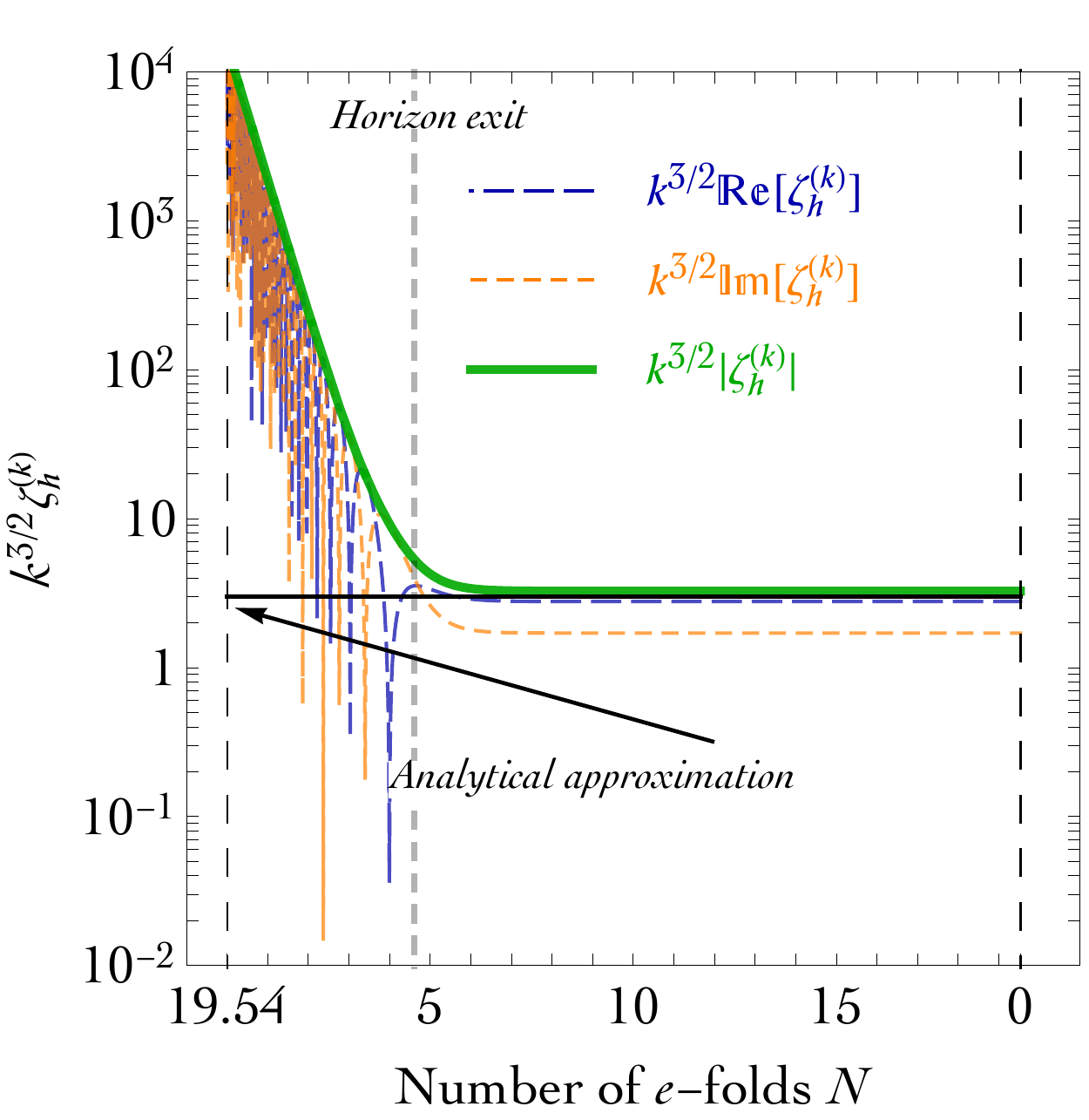}$$
\caption{\label{fig:HiggsEvolution}\em 
{\bf Left}: Sample evolution of the classical Higgs background ($h_{\rm cl}$, red solid line) and of a perturbation with $k/a_{\rm in}H_0 = 10^2$ (dashed lines).
{\bf Right}: 
Higgs curvature perturbation $\zeta_h^{(k)}$ during inflation.
We compare the full numerical result 
with the analytical approximation (last term in eq.~(\ref{eq:CurvatureDuringInflation}), 
 solid horizontal black line). The vertical dashed gray line marks the instant of horizon exit.
We use the analytical approximation in eq.~(\ref{eq:Potential}) with  $h_{\rm cr}=4~10^{12}\GeV$,
$b=0.09/(4\pi)^2$ (which corresponds to $M_t = 172$ GeV) and $H_0=10^{12}\GeV$.
}
\end{center}
\end{figure}

We are now in the position to solve eq.~(\ref{eq:MS2}). To this end, we need boundary conditions for $u_k$ and its time derivative. 
 We use the Bunch-Davies conditions at $N = N_{\rm in}$ for 
 modes that are sub-horizon at the beginning of the classical evolution, $k/a_{\rm in}H_0 \geqslant 1$,
and we treat the real and imaginary part of $u_k$ separately since they behave like two independent harmonic oscillators for each comoving wavenumber $k$.
At generic $e$-fold time $N$, the perturbation $\delta h_k$ is related to the Mukhanov-Sasaki variable $u_k$ by 
\begin{equation}
\left.k^{3/2}\frac{\delta h_k}{H_0}\right|_{N} = \left.\left(
\frac{k}{a_{\rm in} H_0}
\right)e^{N - N_{\rm in}}(\sqrt{k}u_k)\right|_{N}~.
\end{equation}
We show in the left panel of fig.~\ref{fig:HiggsEvolution} our results for the time evolution of the classical background $h_{\rm cl}$ and the perturbation $\delta h_k$ (both real and imaginary part) during the last $20$ $e$-folds of inflation. As a benchmark value, we consider an initial sub-horizon mode with $k/a_{\rm in}H_0 = 100$. 
After few $e$-folds of inflation  such mode exits the horizon:
oscillations stop, and from this point on, further evolution is driven by the time derivative of the classical background. 
This is a trivial consequence of the equations of motion on super-horizon scales. 
Differentiating eq.~(\ref{eq:ClassicalMotion}) with respect to the cosmic time shows that
$\dot{h}_{\rm cl}$ and $\delta h_k$ satisfy the same equation on 
 super-horizon scales, and, therefore, they must be proportional, $\delta h_k = C(k)\dot{h}_{\rm cl}$ 
 for $k \ll aH_0$~\cite{Espinosa:2017sgp}. 
 The proportionality function $C(k)$ can be obtained by a matching procedure. 
 Deep inside the horizon, in the limit $k \gg aH_0$,  
 the Mukhanov-Sasaki variable $u_k \equiv a\, \delta h_k$ 
   reproduces the preferred vacuum of an harmonic oscillator in flat Minkowski space, and 
    we have, after introducing the conformal time $\tau$ as $dt = a\, d\tau$,  $u_k = e^{-ik\tau}/\sqrt{2k}$. 
    Roughly matching the absolute value of the solutions at horizon crossing  we determine the absolute value of $C(k)$ as
    \begin{equation}\label{eq:SuperHubbleApprox}
   |\delta h_k| = \frac{H_0}{\sqrt{2k^3}\dot{h}_{\rm cl}(t_k)} \dot{h}_{\rm cl}(t)~,
    \end{equation}
where we indicate with $t_k$ the time of horizon exit for the mode $k$ ---  the time
 at which  $k = a(t_k)H_0\equiv a_k H_0$  (equivalently, $-k\tau_k = 1$). The number of $e$-fold at horizon exit is given by 
\begin{equation}
\frac{k}{a_{\rm in} H_0} = e^{N_{\rm in} - N_k}~.
\end{equation}
\subsubsection*{Primordial curvature perturbations}

 The primordial curvature perturbation $\zeta(\vec{x},t)$ on uniform energy density slices $\rho$ is defined (at the linear order) by the perturbed line element~\cite{Bardeen:1983qw,Wands:2000dp} 
\begin{equation}
ds^2 = a^2(t)[1+2\zeta(\vec{x},t)]\delta_{ij}dx^i dx^j~,
\end{equation}
and it is related to the (total) energy density perturbation $\delta\rho$ and to the curvature 
perturbation on a generic slicing $\Psi$ by the gauge invariant formula 
\begin{equation}
\zeta = \Psi + \mathcal{H}\frac{\delta\rho}{\rho^{\prime}} = \Psi + H\frac{\delta\rho}{\dot{\rho}}~,
\end{equation}
where we introduced the conformal time by means of 
$dt = a\,d\tau$ and $\mathcal{H} = aH$ where $\mathcal{H} = a^{\prime}/a$, and prime $^{\prime}$ indicates derivative with respect to $\tau$.
The virtue of this definition is that, 
on super-horizon scales, $\zeta$ is practically identical to the comoving curvature perturbation $\mathcal{R}$
defined on hypersurfaces of constant comoving time. 
Furthermore, on super-horizon scales $\zeta$ is conserved provided that the pressure perturbation is adiabatic.
In conventional single-field inflation models, purely adiabatic perturbations are generated due to quantum fluctuations of the single field driving inflation. 
The considered setup differs from a conventional scenario because of the presence, in addition to the inflaton, 
 of the Higgs field~\cite{Espinosa:2017sgp}.
During inflation,  the curvature perturbation on uniform energy density slices is given, in spatially-flat gauge, 
by
\begin{equation}
\zeta = H_0\frac{\delta\rho}{\dot{\rho}} = \frac{\dot{\rho}_{\phi}}{\dot{\rho}}\zeta_{\phi} + 
 \frac{\dot{\rho}_{h}}{\dot{\rho}}\zeta_{h}~,
\end{equation}
where we separated the inflaton component $\zeta_{\phi}$ 
and the Higgs component $\zeta_h \equiv H_0\delta\rho_h/\dot{\rho}_h$~\cite{Espinosa:2017sgp}. 
We assume that there is no energy transfer between the Higgs and the inflaton sector, and that the latter generates, on large scales, the perturbations  responsible for the CMB anisotropies and large scale structures.
Given these assumptions, $\zeta_{\phi}$ and $\zeta_{h}$ are separately conserved on super-horizon scales~\cite{Malik:2004tf}.
As customary, we can decompose $\zeta_h$ in Fourier modes.
 For a given comoving wavenumber $k$ we have, in spatially-flat gauge
\begin{equation}\label{eq:CurvatureDuringInflation}
\zeta_h^{(k)} = H_0\frac{\delta \rho_h^{(k)}}{\dot{\rho}_h}  = -\frac{\left[\dot{h}_{\rm cl} \dot{\delta h}_k + V_{\rm eff}^{\prime}(h_{\rm cl})\delta h_k\right]}{3\dot{h}_{\rm cl}^2} \simeq \left.
\frac{H_0^2}{\sqrt{2k^3}\dot{h}_{\rm cl}(t_k)}
\right|_{k \ll  aH_0}~,
\end{equation}
where the last analytical approximation is valid for the absolute value of $\zeta_h^{(k)}$ in the 
super-horizon limit, and where we used $\rho_h = \dot{h}_{\rm cl}^2/2  + V_{\rm eff}(h_{\rm cl})$ 
 in the first equality.  

\smallskip 
 
The right panel of fig.~\ref{fig:HiggsEvolution} compares  the 
numerical result for $\zeta_h^{(k)}$ --- that is the first equality in eq.~(\ref{eq:CurvatureDuringInflation}) --- with its analytical approximation. 
The plot, moreover, shows that on super-horizon scales $\zeta_h^{(k)}$ stays 
constant, as it should be since we are working under the underlying assumption that there are no interactions between the Higgs and the inflationary sector.

\subsection{Higgs dynamics after inflation}\label{postinf}
Assuming instantaneous reheating, the energy density of the inflaton 
is instantaneously converted into radiation at the end of inflation.
The inflaton energy density $\rho_{\rm R}$ is related to the temperature of the thermal bath  by 
\begin{equation}\label{eq:EnergyRad}
\rho_{\rm R} = \frac{\pi^2 g_*}{30}\, T^4~.
\end{equation}
The dynamics after reheating is described by the following system of coupled Higgs-radiation equations
\begin{eqnsystem}{sys:HR}
\ddot{h}_{\rm cl} + \left(3H + \Gamma\right)\dot{h}_{\rm cl} + V^{\prime}_{\rm T}(h_{\rm cl}) &=& 0~,
\label{eq:ThermalHiggs}\\
\dot{\rho}_{\rm R} + 4H\rho_{\rm R} &=& \Gamma\rho_{h}~,\label{eq:ThermalRad}
\end{eqnsystem}
where the energy density of the Higgs field is given by 
\begin{equation}\label{eq:HiggsEnergyDensity}
\rho_{h} = \frac{\dot{h}_{\rm cl}^2}{2} + V^{\rm T}_{\rm eff}(h_{\rm cl})~,
\end{equation}
and the Hubble parameter is related to the total energy density by 
\begin{equation}\label{eq:HubbleAfterReh}
H^2 = \frac{\rho_{\rm R} + \rho_{h}}{3\bar{M}_{\rm Pl}^2}~.
\end{equation}
The damping factor $\Gamma$ takes into account the Higgs decays at temperature $T$, and represents the energy transportation from the Higgs field to radiation. We use the expression quoted in~\cite{Espinosa:2017sgp}, $\Gamma \simeq 10^{-3}T$. 
In the following, we shall adopt the assumption of sudden decay approximation. In this approximation --- corresponding to $\Gamma = 0$ in eq.s~(\ref{sys:HR}) --- Higgs and radiation evolve separately, and  
Higgs decay occurs 
 instantaneously at $H = \Gamma$.  The system in eq.s~(\ref{sys:HR}) can be solved using the following boundary conditions. As far as the classical Higgs field is concerned, we use the field values at the end of inflation computed in section~\ref{sec:BeforeInf}. The evolution of $\rho_{\rm R}$, on the contrary,
 starts from the temperature in eq.~(\ref{eq:ReheatingT}).\footnote{More precisely
 conservation of total energy determines the reheating temperature  as 
 \begin{equation}
 3\bar{M}_{\rm Pl}^2H_0^2 + \left.\frac{\dot{h}_{\rm cl}^2}{2}+ V_{\rm eff}(h_{\rm cl})\right|_{h_{\rm cl} = h_{\rm end}} = 
 \frac{\pi^2 g_*}{30}\, T^4 + 
 \left.\frac{\dot{h}_{\rm cl}^2}{2}+ V_{\rm eff}^{\rm T}(h_{\rm cl})\right|_{h_{\rm cl} = h_{\rm end}}.
 \end{equation}
The approximation in eq.~(\ref{eq:ReheatingT}) is valid because
the inflaton energy density dominates over the Higgs contribution.}
 Using the solution for $\rho_{\rm R}$, eq.~(\ref{eq:EnergyRad}) gives the time-evolution of the temperature.
 
After instantaneous reheating, the Higgs potential suddenly changes   
 from the expression in eq.~(\ref{eq:Potential}) to the one in eq.~(\ref{eq:ThermalPot}), 
 and interactions with the SM bath generate a large thermal mass. 
 If $h_{\rm end} < h_{\rm max}^{\rm T}$, 
 the fall of the Higgs 
 field is rescued by thermal corrections, and its background value starts oscillating around the 
 minimum at zero until the decay becomes efficient at $H = \Gamma$. 
 At this stage, the Higgs component $\zeta_h$ of the curvature perturbation is not conserved compared to the value 
 computed in eq.~(\ref{eq:CurvatureDuringInflation}) during inflation. 
 This is because the interactions with the SM plasma that are responsible for 
  the appearance of the thermal mass introduce a non-adiabatic component 
 in the pressure perturbation.
 
 We compute --- in spatially flat gauge, and for a given comoving wavenumber $k$  ---
 the total curvature perturbation after reheating according to 
 \begin{equation}\label{eq:CurvatureAfterInflation}
\zeta^{(k)} = H\frac{\delta \rho}{\dot{\rho}}   =
\frac{\dot{h}_{\rm cl} \dot{\delta h}_k + V^{\rm T\,\prime}_{\rm eff}(h_{\rm cl})\delta h_k}
{-4\rho_{\rm R} - 3\dot{h}_{\rm cl}^2}~,
\end{equation}
where radiation perturbations are set to zero. We compute $\zeta^{(k)}$ numerically 
at the time of Higgs decay, $H = \Gamma$.  
After Higgs decay, radiation remains as the only component of the Universe
and $\zeta^{(k)}$ is, therefore, conserved.

\begin{figure}[t]
\begin{center}
\includegraphics[width=.45\textwidth]{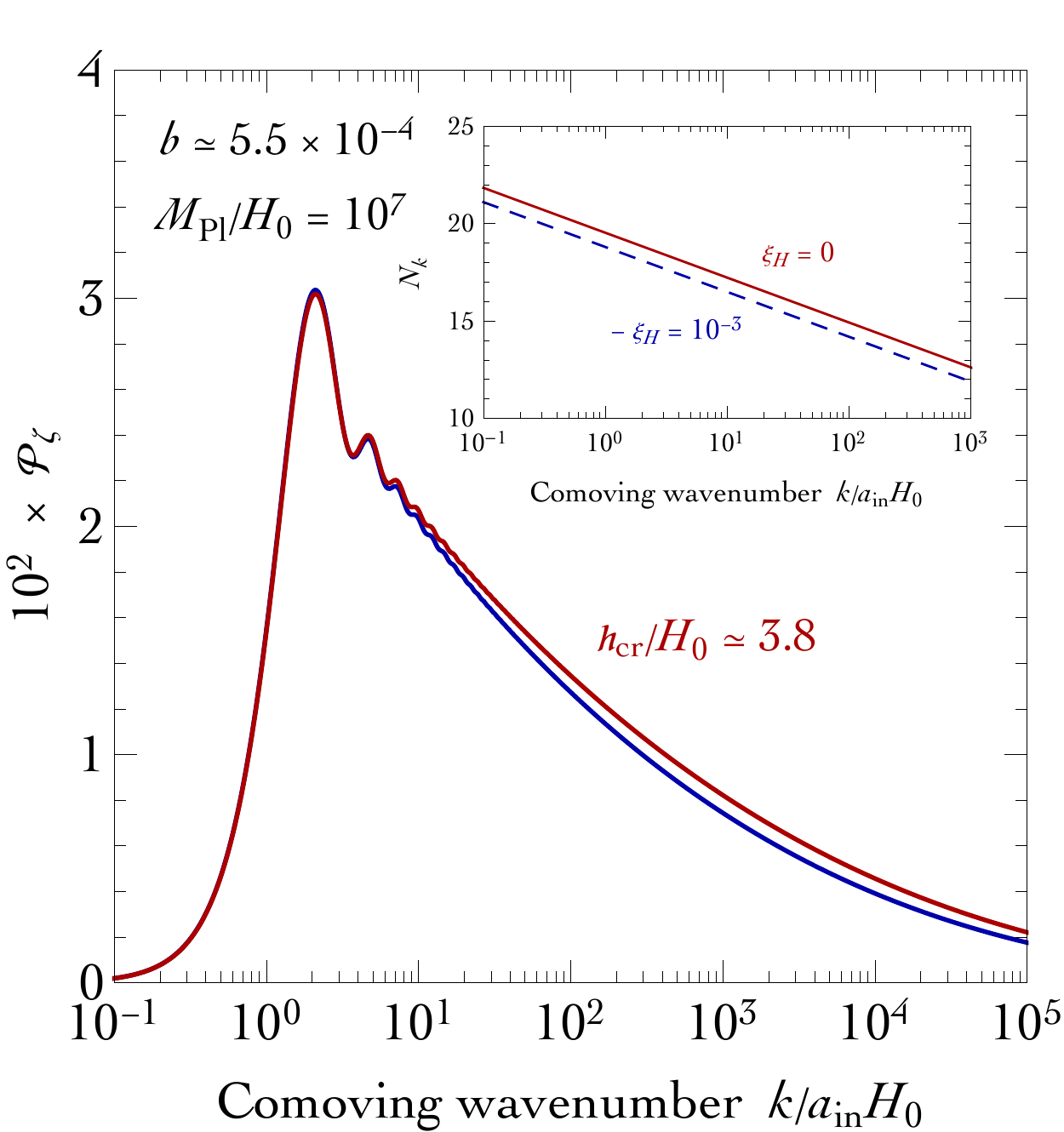}\qquad
\includegraphics[width=.45\textwidth]{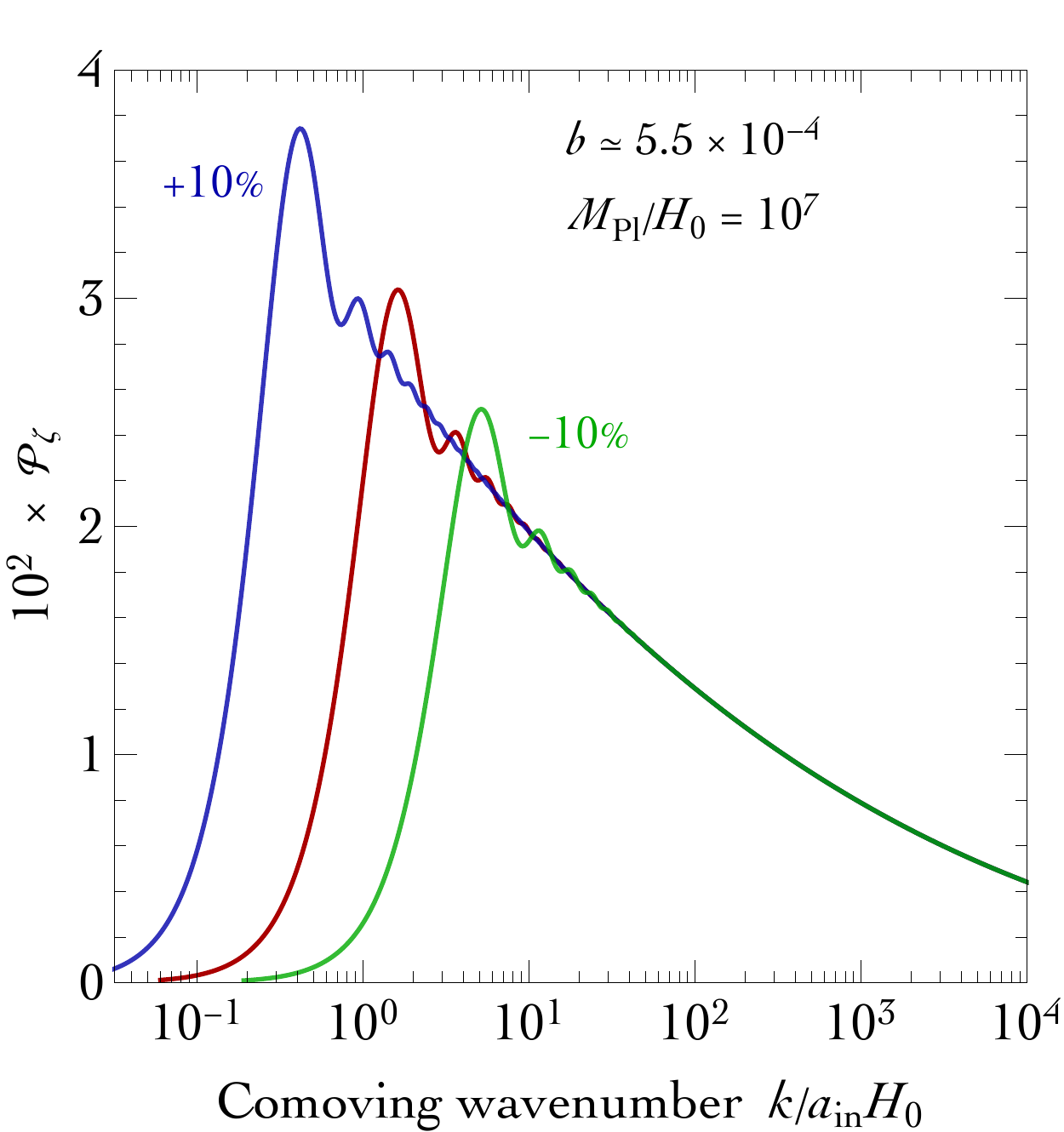}
\caption{\em\label{fig:PowerSpectrum} {\bf Left:} power spectrum of Higgs fluctuations produced by rescued Higgs fall
for $\xi_H=0$ (red curve) and $\xi_H=-10^{-3}$ (blue), which has a minor impact. 
In the inset we show the number of $e$-folds at horizon exit for each mode.
{\bf Right}: how the power spectrum changes when the classicality condition in eq.\eq{ClassicalvsQuantum}
is changed by $\pm 10\%$.
}
\end{center}
\end{figure}

\subsection{The power spectrum and the PBHs abundance}\label{BH}
The curvature power spectrum is given by 
\begin{equation}
\mathcal{P}_{\zeta} = \frac{k^3}{2\pi^2}|\zeta^{(k)}|^2~.
\end{equation}
A numerical example is shown in fig.\fig{PowerSpectrum}.
Its left panel shows that a small $\xi_H = -10^{-3}$ only has a minor effect.
The cut in the power spectrum at small $k$ arises because of assumption $i)$:
before that classical rolling starts,
the Higgs field is away from its minimum and exactly homogeneous.
Relaxing this assumption would increase the power spectrum at small $k$.

The right panel shows the significant effect of a small change in the arbitrary order one parameter $c$
that defines the classicality condition in eq.\eq{ClassicalvsQuantum}.
Reducing $c$ anticipates the initial moment $t_{\rm in}$ (or equivalently $N_{\rm in}$)
where the Higgs starts to roll down the potential.
As a consequence earlier quantum fluctuations get taken into account by our computation
of section~\ref{sec:BeforeInf}.

\medskip

Finally, we can now compute the mass and amount of PBH generated by Higgs fluctuations.
The radius of hubble horizon or the wavelengh of the modes determines the typical mass of the PBHs~\cite{astro-ph/9901293}:
\begin{equation}
   M_{\rm PBH} \approx \frac{ \gamma }{2}\frac{ M_{\rm Pl}^2 } {H_0} e^{2 N}  \ , 
\end{equation}
where $N$ is the number of $e$-folds when the $k$-mode leave the horizon;
$\gamma \approx 0.2 $ is a correction factor~\cite{Carr:1975qj}. 
For example $H_0 = 10^{12} \,\mathrm{GeV} $ and $N= 20$ gives $M_{\rm PBH}\approx 10^{-15} M_\odot$.

To compute the fraction of the Hubble volume collapsing to PBHs, we need  the variance of the 
smoothed density perturbation over a radius $R$,
$   \sigma^2(R) = \langle \delta^2 ( \mb{x}, R ) \rangle$,
where $\delta ( \mb{x}, R ) = \int \mathrm{d}^3{x}^\prime  \, \delta ( \mb{x}' ) 
         W ( \mb{x} - \mb{x}' , R )$
is the density fluctuation  smoothed by a window function  $W ( \mb{x} , R )$, assumed to be Gaussian
\beq W ( \mb{x} , R ) = 
\frac{1} { (2 \pi)^{3/2}R^3 } \exp \left( - \frac{ |\mb{x}|^2}{2 R^2}\right).\eeq
The variance can be computed in terms of density power spectrum $P_\delta (k ) $,
which is related  to the curvature perturbation $\zeta$ power spectrum $P_\zeta(k)$ as~\cite{1801.05235} 
\begin{equation}
   \sigma^2(R) = \int \mathrm{ d } \ln k  \, P_\delta (k) \,\widetilde{W}^2 ( kR )   , 
      = \int \mathrm{ d } \ln k  \frac{16}{81}( kR )^4 P_\zeta (k) \widetilde{W}^2 ( kR )   , 
\end{equation}
where $ \widetilde{W} ( k ) = \exp \left( - \sfrac{k^2}{2}\right)$ is the Fourier transform of the window 
function. 
Assuming that PHBs are  formed when
 the density perturbation exceeds $\delta_{\rm th} \approx 0.5 $~\cite{gr-qc/0412063},
the fraction of the Universe ending up in PHBs is given by the tail of the assumed Gaussian  distribution: 
\begin{eqnarray} 
\beta ( M_{\rm PBH} ) &=& \gamma \int^\infty_{\delta_{\rm th}} \mathrm{d} \delta \frac{ 1}{ \sqrt{2 \pi} \sigma( M_{\rm PBH} )  }
      \exp \left( -\frac{  \delta^2} { 2 \sigma^2 (M_{\rm PBH} )   } \right)  \stackrel{\sigma \ll \delta_{\rm th}}{\simeq}
        \frac{\gamma \sigma}{  \sqrt{2 \pi}  \delta_{\rm th} }\exp \left(  - \frac{\delta_{\rm th}^2} { 2\sigma^2} \right)      \ , 
   \label{eq:betaM}
\end{eqnarray} 
The latter approximation is relevant, given that the obtained power spectra $P_\zeta$ are of order $10^{-2}$.
We convert $\sigma(R)$ to a function  $\sigma(M_{\rm PBH})$, taking into account that
the size $R$ is related to the mass $M_{\rm PBH}$ as $R \approx 2 G M_{\rm PBH}/ \gamma a$.
The fraction of PHB relative to the DM abundance at given mass, $f_{\rm PBH} ( M_{\rm PBH})$, is given by~\cite{1801.05235}
\begin{equation}
   f_{\rm PBH} \approx  2.7 \times 10^{8} 
         \bigg( \frac{ 10.75}{ g_{*,\rm form}} \bigg)^{1/4} 
         \bigg(\frac{\gamma}{0.2} \frac{ M_\odot}{M_{\rm PBH}}  \bigg)^{1/2} 
         \beta   \ . 
\end{equation}
The distribution of PHBs as function of their mass $M_{\rm PBH}$ is strongly peaked at the value that maximises $\sigma(M)$, in view of
the exponential factor in eq.~(\ref{eq:betaM}).
In terms of the power spectrum $P_\zeta$  this means that 
the PBH abundance roughly scales as $e^{-1/P_\zeta}$ and 
is dominated by the top of the peak of $P_\zeta(k)$.
The PBH mass distribution is peaked at the value corresponding to the $k$ that maximises $P_\zeta(k)$.
The desired abundance is reproduced for $\max P_\zeta \sim 10^{-2}$, and slightly larger (smaller) 
values produce way too many (too few) PBH.

This means that, due to the fine-tuned nature of the mechanism
(as in all models that can generate PBH)
the amount and mass of PBH depends in an extremely sensitive way on the uncertain SM
and cosmological parameters, mainly the top mass and the Hubble constant $H_0$.
Furthermore, uncertainties in the computation of black hole formation (such as the value of $\delta_{\rm th}$)
imply uncertainties of many order of magnitude in the PBH density.

For example, a $10\%$ variation in the order one arbitrary constant $c$ that parameterizes the classicality condition has a order one impact on the power spectrum (fig.\fig{PowerSpectrum}), 
and consequently a huge impact on the PBH abundance.

In addition to SM parameters and to $H_0$, the
PBH abundance also depends on two extra parameters 
that characterise the assumed initial condition:
constant $h(x) =h_{\rm in}$ at $N_{\rm in}$ $e$-folds before the end of inflation,
when the approximated classicality condition is satisfied.
The next section reconsiders if this assumption is justified.


\subsection{Is a homogeneous Higgs background a sensible assumption?} \label{Homo}
The computation was based on assumption $i$):
at $N_{\rm in}\approx 20$ $e$-folds before the end of inflation, the
Higgs field must be away from its minimum and constant within the presently observable horizon.

 \begin{figure}[t]
\begin{center}
$$\includegraphics[width=.45\textwidth]{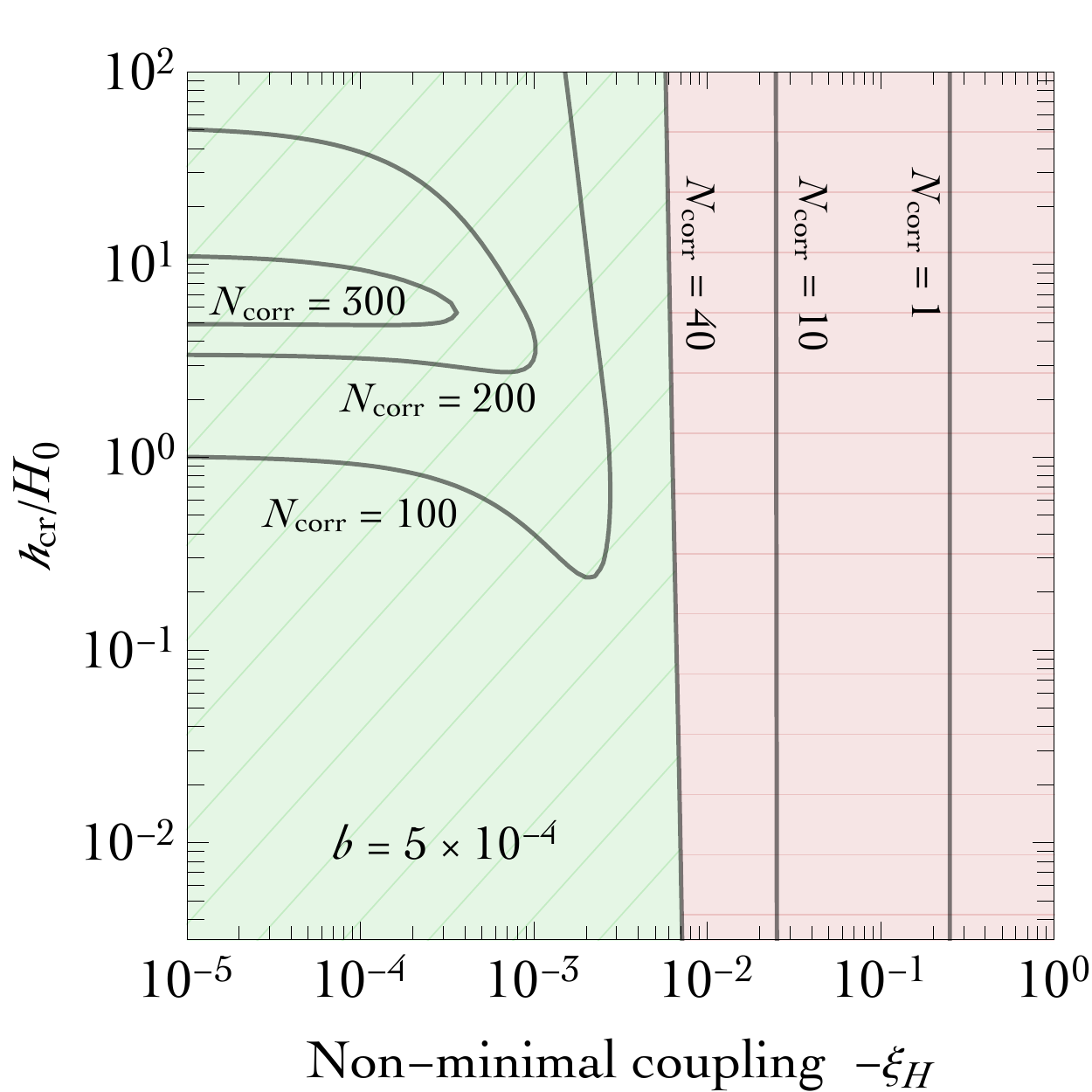}$$
\caption{\em \label{fig:Corr2} 
Contour plot of the correlation length $N_{\rm corr}$
of Higgs inflationary fluctuations as function of the main parameters, $\xi_H$ and $h_{\rm cr}/H_0$.
The needed homogeneous patch, $N_{\rm corr} > 40$, can be obtained in the green region, 
which corresponds to a small negative $\xi_H$.
}
\end{center}
\end{figure}

\subsubsection*{Approximate Higgs homogeneity?}
We here show that approximate homogeneity is a natural product of inflation, provided that the SM Higgs potential satisfies certain conditions.
Quantum corrections in inflationary (de Sitter) space-time have been studied in~\cite{Starobinsky:1994bd}, that showed that
long wavelength fluctuations can be described by a Fokker-Planck equation
for $\rho(h,N)$, the probability of finding the Higgs field at the value $h$ at $N$ $e$-folds of inflation:
\beq \label{eq:evoN}
-\frac{\partial\rho}{\partial N} = \frac{H_0^2}{8\pi^2}\frac{\partial^2\rho}{\partial h^2}
 + \frac{\partial}{\partial h}\left(\frac{V_{\rm eff}^{\prime}\rho}{3H_0^2}\right)~.
\eeq
The first term on the right-hand side is a diffusion term due to quantum fluctuations while
 the second term is a drift (or transport) term due to the potential.
After some $e$-folds, the distribution converges to its
 equilibrium  value  $\rho_{\rm eq}(h) \propto \exp(-8\pi^2 V_{\rm eff}(h)/3H^4)$~\cite{Starobinsky:1994bd}.
 
We are interested in the probability of having a roughly constant Higgs away from its minimum $h=0$
within the presently observable horizon.
This configuration is a natural outcome of inflationary dynamics provided that the correlation length of Higgs fluctuations is larger
that the present horizon.
Following~\cite{Starobinsky:1994bd}, the computation of the correlation length 
is simplified observing that correlation functions depend on space and time separations
respecting the O(4,1) invariance of de Sitter.
Thereby we can  compute the evolution of Higgs fluctuations at a fixed point in space,
and study the correlation as function of time $t$, 
or equivalently as function of the number of $e$-folds $N$.
At large time separation the correlation is well approximated by its dominant exponential,
and parameterized in terms of a correlation time $\tau_{\rm corr}$
or equivalently in terms of number of $e$-folds  $N_{\rm corr} = H_0 \tau_{\rm corr}$ as~\cite{Starobinsky:1994bd}
\beq \label{eq:CorrelationLen}
\med{h(t_1, \vec x) h(t_2,\vec x)} \simeq  \med{h^2} e^{-|t_1-t_2|/\tau_{\rm corr}} =\med{h^2} e^{-|N_1-N_2|/N_{\rm corr}}. 
\eeq
O(4,1) invariance implies that the spatial correlation length is exponentially large $e^{H N_{\rm corr}}$,
namely space is exponentially inflated.
We demand that $N_{\rm corr}$ is larger than about 40 in order to produce a smooth region as large as our Universe
at $N_{\rm in}\approx 20$ $e$-folds before the end of inflation.
Before computing numerically $N_{\rm corr}$ for the SM potential, it is useful to consider some simple limits:
\begin{itemize}

\item[0)]  
A massless free scalar $h$ is a simple but
special case, because it does not `thermalise' to the equilibrium distribution.
Rather, it undergoes  random walk diffusion.
Starting from $h=0$, 
after $N$ $e$-folds one has ${\langle  h^2\rangle}^{1/2}= \sqrt{N} H_0/2\pi$
and the correlation length in a region with given $\langle h\rangle$ is 
$N_{\rm corr} = 2\pi  \med{h}/H $.
Imposing $N_{\rm corr} > 40$ implies $H_0< 0.16 \med{h}$.

\item[$m)$] 
A  free scalar with squared mass $m^2>0$ fluctuates as 
${\langle  h^2\rangle}{}^{1/2}=\sqrt{3/2}H^2/2\pi m  $
with correlation length $N_{\rm corr} = 3H^2/m^2$~\cite{Starobinsky:1994bd}.
For $m^2 = - 12\xi_H H^2$ one gets $N_{\rm corr} = -1/4\xi_H > 40$ for $ -0.006<\xi_H <0$.
Roughly, this will be our final result.

\item[$\lambda)$] 
A massless scalar with an interaction $\lambda h^4/4$ and $\lambda>0$ fluctuates as 
${\langle  h^2\rangle}{}^{1/2}= \sfrac{0.36 H}{ \lambda^{1/4}}$
with $N_{\rm corr} = 7.62/\sqrt{\lambda}$.
So $N_{\rm corr} > 40$ for $\lambda < 0.036$.
This is satisfied in the SM at large energy. 
\end{itemize}
The above two results agree, up to order one factors, 
approximating a generic $V(h)$ as a massive field with $m^2 = V''$.
The precise value of the correlation length can be obtained as proposed in~\cite{Starobinsky:1994bd}:
the dominant exponential that solves eq.\eq{evoN} can be computed as the eigenvalue of a Schroedinger-like equation.
{In the SM $\lambda_{\rm eff}(h)$ runs to negative values, making the curvature $V''$ negative at large field values; 
the patch with a large correlation length is created while $h$ is climbing the potential in its region with positive curvature.}
The result is shown in fig.\fig{Corr2}a, 
and basically agrees with the result anticipated at point $m)$:
$\xi_H$ must be negative and small. Such a small $\xi_H$ is roughly compatible with the size of quantum corrections to $\xi_H$.
The running Higgs quartic coupling is small enough that it does not spoil the mechanism, as qualitatively understood
at point $\lambda)$.

In conclusion, 
a roughly homogeneous Higgs field (up to fluctuation of order $H_0$) encompassing our whole horizon
is a natural outcome of inflation, provided that $\xi_H$ is small enough.
In a multiverse context, its fine-tuned value that leads to DM as PBH can be justified on anthropic basis.

 \begin{figure}[t]
\begin{center}
$$\includegraphics[width=.45\textwidth]{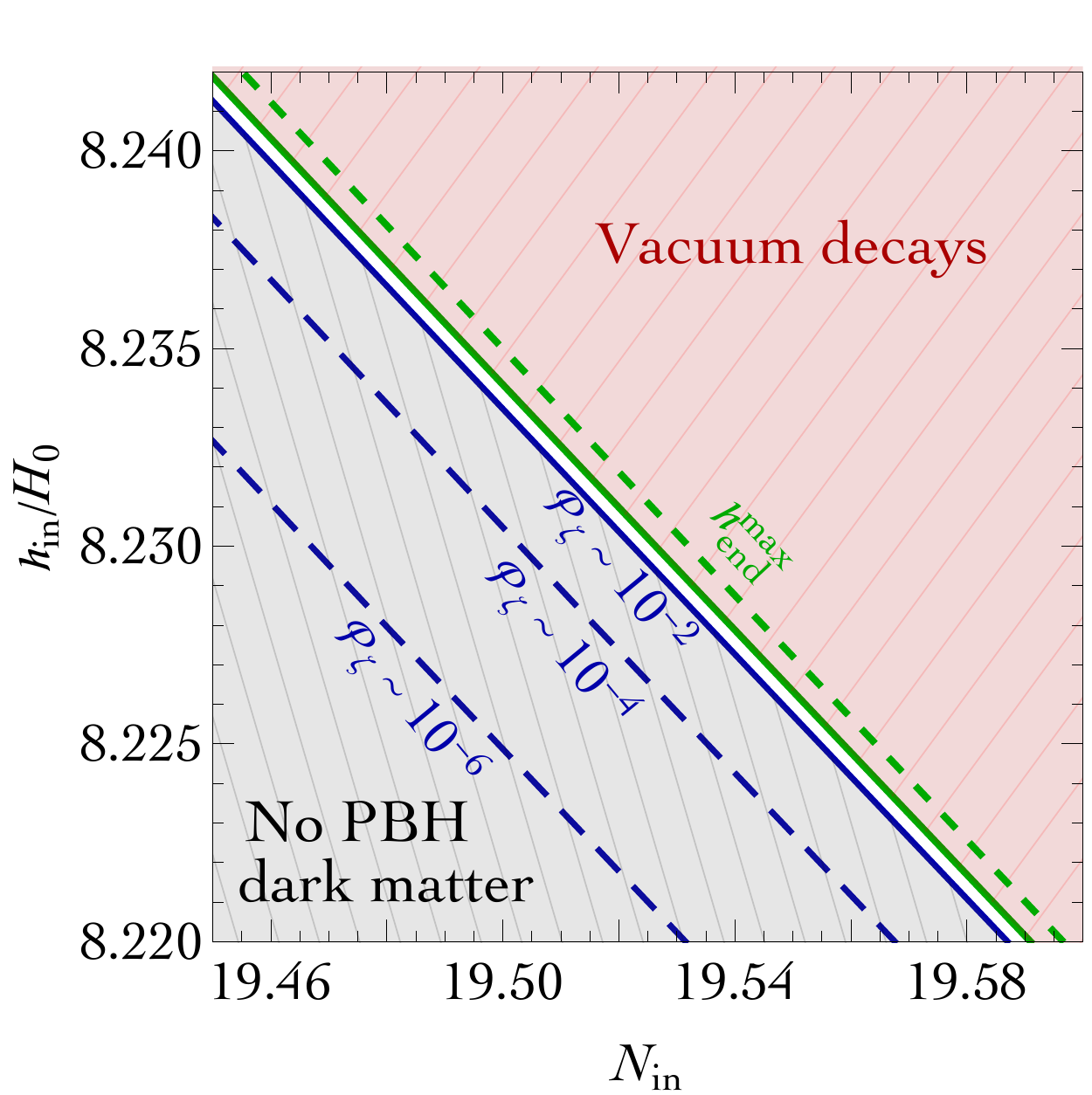}\qquad
\includegraphics[width=.465\textwidth]{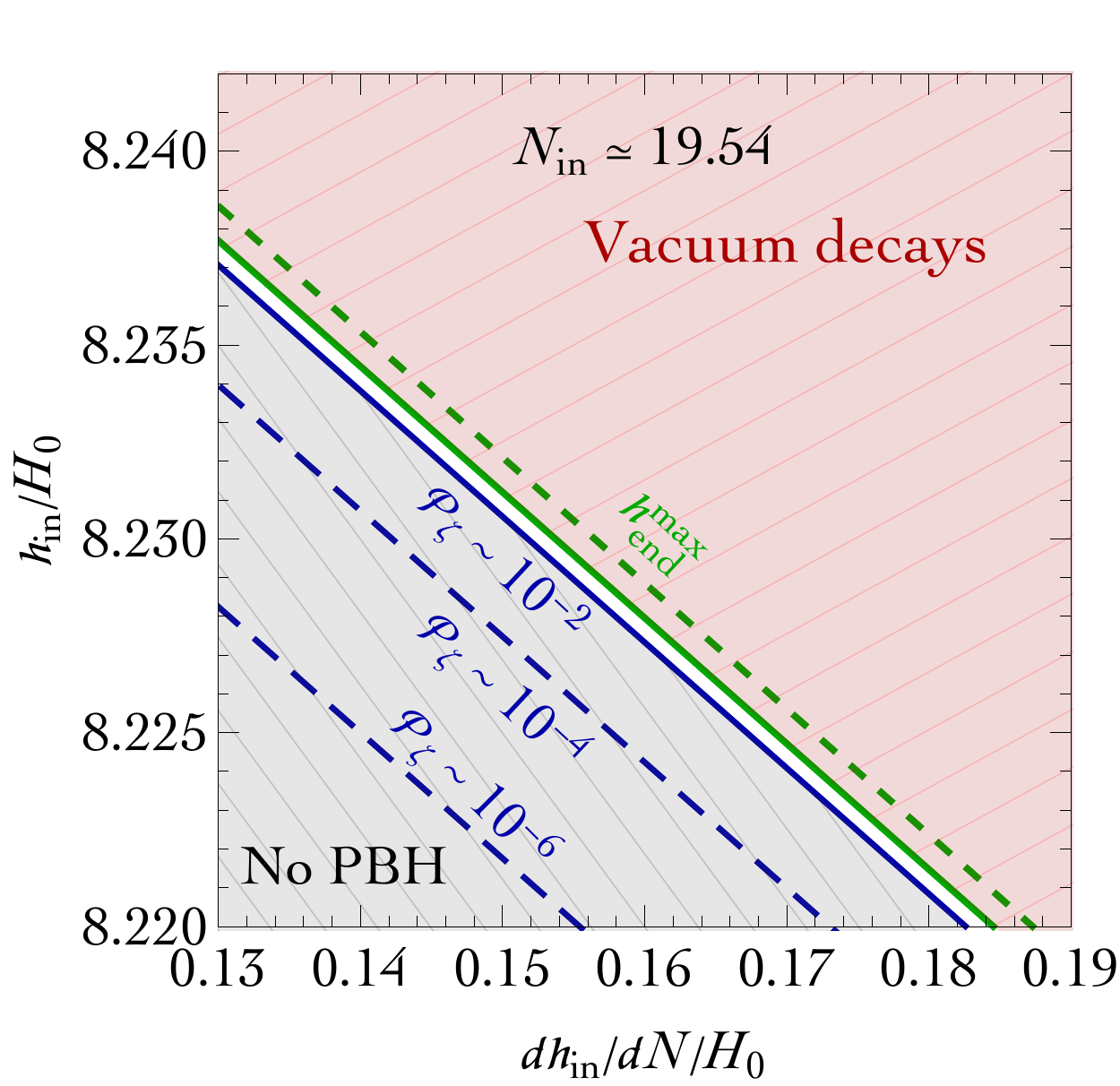}$$
\caption{\em \label{fig:ahi} 
We fix the SM to $M_t=172\GeV$ (which corresponds to $h_{\rm cr}=4~10^{12}\GeV$,
$b=0.09/(4\pi)^2$) and $H_0=10^{12}\GeV$
and explore the dependence on the parameters that define the assumed initial condition.
{\bf Left}: Final outcome as function of the homogeneous initial value of the Higgs field $h_{\rm in}$ 
at $N_{\rm in}$ $e$-folds before the end of inflation and of $N_{\rm in}$.
The desired PBH abundance is produced inside the white band.
Inflationary fluctuations in $h_{\rm in}$ are much larger than the needed tuning.
{\bf Right}: We fix $N_{\rm in}$, and introduce $dh_{\rm in}/dN$ as additional parameter.
}
\end{center}
\end{figure}

\subsubsection*{Exact Higgs homogeneity?}
Fig.\fig{ahi}a shows that the initial homogeneous Higgs fields $h(x)=h_{\rm in}$ at $N=N_{\rm in}$ $e$-folds before the end of inflation must be tuned  to a part in about $10^{-3}$
in order to produce a final $h_{\rm end}$ close to the maximal value that can be rescued by reheating,
as needed to produce a substantial amount of primordial black holes.
An interesting PBH abundance is obtained within the
narrow strip in fig.\fig{ahi}a. Its boundaries have been computed as follows:
\begin{itemize}
\item A slightly smaller $h_{\rm in}$ leads to a negligible PBH amount.
In fig.\fig{ahi}a we show the peak value of the power spectrum, and shade in gray regions where it is below $10^{-2}$.


\item 
A  slightly larger $h_{\rm in}$ leads to a too large $h_{\rm end}$ not rescued by reheating.
In fig.\fig{ahi}a we shade in red regions where $h_{\rm end}$ is above the maximal value 
that can be rescued by instantaneous reheating.

\end{itemize}
In fig.\fig{ahi}b we show that a variation in $\dot h_{\rm in}$ has a smaller effect, as it gets red-shifted away.

Within the assumption that $h_{\rm in}$ is homogeneous, its tuned value can be justified on anthropic basis
provided that PBH make all DM~\cite{DMant,Espinosa:2017sgp}.
However, one single fluctuation $\delta h_{\rm in}\circa{>}10^{-3}  h_{\rm in}$ away
from the assumed perfect
homogeneity can lead to one vacuum decay bubble that, after inflation, 
expands engulfing the observable Universe.

As discussed previously, inflation can produce an approximate homogeneity within a large patch,
but up to fluctuations of order $\delta h_{\rm in}\sim H_0/2\pi$.\footnote{Technically, such fluctuations appear in our equations when (motivated for example by the arbitrariness in the classicality condition)
we start from a different initial time: this shifts the position and the height of the presumed peak of the power spectrum, and thereby the  abundance and mass of PBHs.}
Additional fluctuations in $\dot h_{\rm in}$ are less significant and we ignore them.
The Hubble constant $H_0$ cannot be significantly reduced, for two reasons.
First, $H_0\circa{>}h_{\rm cr}$ is needed to
keep the Higgs fluctuating around the top of its potential barrier,
until it starts to classically roll down.
Second, it would suppress the unwanted pre-fall fluctuations $\delta h_{\rm in}$
at the price of suppressing post-fall fluctuations that generate PBHs, see eq.\eq{CurvatureDuringInflation}.

Thereby we must take into account the effect of fluctuations in the initial $h(x)$.
Pictorially, the status of the Universe should now be represented in fig.\fig{Corr2}b 
not by a point (which can lie in the desired region), but by a dot, much thicker than the desired region.

This is a problem because, at $N_{\rm in}\approx 20$ $e$-folds before the end of inflation (which needs at least $60$ $e$-folds),
the present Universe is composed by about $ e^{120}$ causally disconnected regions.
Within each region, the probability that the Higgs fluctuates to the desired tuned $h_{\rm in}$ is  about $10^{-2-3}$.

Under-fluctuations produce a negligible PBH abundance.
On small scales this is not a problem: only the average matters.
Accretion of black holes after their formation would increase the average.
On large scales of the order of the present horizon
(those probed by observations, that find a DM density 
more homogeneous than what could be justified anthropically),
fluctuations of $h(x)$ would produce a PBH density which is not homogeneous.

\smallskip

The effect of over-fluctuations is much worse.
Like in a cosmic russian roulette, 
a too large $h_{\rm in}$ in one of the $e^{120}$ regions can form a
vacuum decay bubble that, after the end of inflation, engulfs the whole Universe
(a general relativity computation finds that innocuous 
bubbles that shrink and/or expand behind a black-hole horizon can form,
but together with dangerous ones~\cite{Espinosa:2017sgp}).
The probability of avoiding vacuum decay is roughly estimated as  $\wp\sim 2^{-e^{120}}$.



\medskip

As we now discuss, this unlikely possibility cannot be justified on anthropic basis.

To explain why, let us start from an analogous anthropic argument considered by Weinberg~\cite{Weinberg:1987dv}:
if observers only exist where the cosmological constant $\Lambda$ is small enough to allow their development
(at the price of a tuning with probability $\wp_\Lambda \sim 10^{-120}$, which is possible
in multiverse with more than $10^{120}$ different vacua), they should expect 
to see a cosmological constant around the anthropic bound.
Once the desired Universe with small $\Lambda$ is formed, it is relatively safe.
Technically, the time-scale for the variation of the $\wp_\Lambda $ is the Hubble scale, presently
$10^{10}\,{\rm yr}$.

Within the `rescued Higgs fall' mechanism, 
the probability $\wp$ to form DM but no vacuum decay is much smaller.
This might be by itself a problem, unless one relies on eternal inflation and argues that $1/\wp$ 
is smaller than infinity.
The same Weinberg-like argument leads to expectations incompatible with experience.
Indeed, the time-scale for the growth of $1/\wp$ is very short.
An observer that justifies its lucky survival to vacuum decay as needed for its existence,
should expect to be immediately executed by an expanding vacuum decay bubble, given that
it arises with statistical certainty in the extra regions which continuously enter in causal contact with the observer
due to the Universe expansion.

\medskip

Needless to say, multiverse probability is a shaky concept, plagued by infinities.
Nevertheless, the problem seems worrying enough that one wonders if it can be avoided or alleviated.

\smallskip

One possibility is devising a mechanism that suppresses vacuum decay by
rescuing the Higgs more efficiently
than the thermal barrier considered in~\cite{1505.04825,Espinosa:2017sgp}.
For example, a non-thermal distribution\footnote{We thank  the authors of~\cite{Espinosa:2017sgp} for
this suggestion.}
or an inflation that initially decays to the SM
particles more coupled to $h$, increasing its thermal mass.
The green dashed line in fig.\fig{ahi} shows the extra rescued region imposing
what we believe is the most optimistic possibility:
$h_{\rm end} <h_{\rm end}^{\rm max}$ with  $V_{\rm eff}(h_{\rm end}^{\rm max}) =0$.
Namely, we demand  that the negative Higgs potential energy remains smaller
than the inflaton potential $V_0 = 3\bar{M}_{\rm Pl}^2H_0^2$, otherwise nothing can stop the Higgs fall.
Fig.\fig{ahi} shows that a more efficient rescue mechanism would not qualitatively change the picture.
Alternatively, some mechanism beyond the SM could prepare the Higgs in the homogeneous state
needed for the `rescued Higgs fall' mechanism.
A study of this possibility goes beyond the scope declared in the title of this paper,
where we wanted to see if some mechanism can generate dark matter within the Standard Model.

\section{Conclusions}\label{conc}
We critically re-examined two different Dark Matter candidates which do not require
new physics beyond the Standard Model.

The first is a hexa-quark $\S=uuddss$ di-baryon, which (being a spin 0 iso-spin singlet)
might have a QCD binding energy large enough to make it lighter than 
$M_\S < 1.876\GeV$ and thereby stable because all possible decay modes are kinematically closed.
In section~\ref{mass} we estimated its mass at the light of recently measured tetra-quarks.
We found that $\S$ could be light enough; possibly as light as $1.2\GeV$, although we cannot provide
a precise mass.
In section~\ref{cosmo} we computed its cosmological relic density, finding that it can reproduce
the desired DM density if $M_\S\approx 1.2\GeV$, while larger masses lead to smaller abundances.
The dominant process that keeps $\S$ in thermal equilibrium 
is scattering of two strange baryons, whose abundance gets Boltzmann suppressed at temperatures
smaller than the strange quark mass, leading to $\S$ decoupling at the temperature
$T_{\rm dec} \approx 25\MeV$ where $\S$ has the desired abundance.
However, in section~\ref{SK} we find, following the strategy of~\cite{Farrar:2003qy}, that such a light $\S$ is excluded, because
nucleons inside nuclei would bind in $\S$ faster than what is allowed by SuperKamiokande
bounds on the stability of Oxygen.
We reached this conclusion at the light of recent global fits of nuclear potentials used to compute
the nuclear wave-function of Oxygen, which indicate that nucleons are close enough to make
$\S$ production too fast.
Both $\S$ and nuclei can be stable for $M_\S\approx 1.87\GeV$; however this mass leads to a
relic $\S$ abundance much smaller than the DM abundance.
Among sparse comments, we mention that direct detection of $\S$ on an anti-matter target gives an annihilation signal.

\medskip

In the second part of the paper, we considered the proposal of~\cite{Espinosa:2017sgp}:
given that the SM Higgs potential can be unstable at large field values,
during inflation the Higgs might fall from an assumed homogeneous vacuum expectation value $h$
beyond the potential barrier towards the Planck scale.
If the fall is tuned such that $h$ almost reaches the maximal value
that can be rescued by thermal effects,
this process generates small-scale inhomogeneities which form primordial black holes.
While DM as PBH seems excluded in the proposed mass range (just above the bound on
Hawking radiation) the proposal is interesting enough to deserve further scrutiny.
We confirmed the computations of~\cite{Espinosa:2017sgp} and extended them
adding a non-minimal coupling $\xi_H$ of the Higgs to gravity.
We find that inflationary fluctuations can generate a quasi-homogenous $h$
only if $\xi_H$ is as small as allowed by quantum corrections that unavoidably generate it.
Furthermore, we find in section~\ref{Homo} that the amount and mass of PBH depend on an extremely sensitive way on the uncertain SM and
cosmological parameters, including two extra parameters introduced as assumption:
the Higgs starts falling from a homogeneous value at a given moment during inflation.
The assumed homogeneity is however not the typical state present during inflation,
where the Higgs has fluctuations of order Hubble. 
This is important for the present mechanism, because it has a Russian roulette feature:
the Universe is eaten by a vacuum decay bubble if 
the Higgs fluctuates to a value too high to be rescued by thermal effects in one of the $e^{120}$
causally disconnected patches in which the present horizon is divided while the Higgs starts falling.
The probability of obtaining the observed Universe is so small, about $(1/2)^{e^{120}}$,
that trying to justify it through anthropic considerations leads to
the issues discussed in section~\ref{Homo}.

\smallskip

We conclude that tentative searches and interpretations
of Dark Matter as a phenomenon beyond the Standard Model remains a justified field.

\appendix

\small

\subsubsection*{Note Added}{G.\ Farrar in \href{https://arxiv.org/abs/1805.03723}{arXiv:1805.03723}
proposed that a co-stable 
$\S$ Dark Matter with $1.86 \GeV < M_\S < 1.88\GeV$ could be produced with roughly the correct relic abundance at the QCD phase transition at $T \approx 150 \MeV$.
We have not considered this possibility, because
this abundance would be washed out by thermal equilibrium
through the $\S$ break-up reactions in eq.\eq{Sprod}.
To avoid this conclusion the $\S$ breaking 
cross sections should be $\sim$ 10 orders of magnitude smaller than the naive QCD estimate,
given by the $\S$ radius squared.
We view this as a too extreme possibility given that ---
while some special suppression could arise at low energy
assuming appropriate nucleon potentials ---
one would need such a strong suppression at $T \approx 150\MeV$,
where the simpler QCD physics is relevant.}

\footnotesize
\bibliographystyle{abbrv}

\subsubsection*{Acknowledgments}
We thank G. Ballesteros, I. Bombaci, A. Bonaccorso,   A. Francis, A. Gnech, J. Green, M. Karliner, D. Racco, Y. Wang for discussions
and M.\ Redi for participating in an early stage of this project.
We especially thank  G.~Farrar for discussions about $\S$ break-up cross sections and
A. Riotto for many useful suggestions at all stages of this work, and in particular for clarifications about non-adiabaticity of $\zeta_h$ because of interactions with the plasma.
A.U. thanks all participants to the Dark Matter Coffee Break 
 at SISSA for many stimulating discussions about the physics of primordial black holes.
 This work was supported by the ERC grant NEO-NAT.

\end{document}